\documentclass[reprint,amsmath,amssymb,pra,floatfix]{revtex4-2}
\pdfoutput=1 \usepackage{graphicx}

\usepackage[colorlinks=true,allcolors=blue]{hyperref}

\DeclareMathOperator{\sinc}{sinc}

\begin{document}

\newcommand{\param}{{\ell}}
\newcommand{\ns}{{\mkern-1mu}}
\newcommand{\ps}{{\mkern1mu}}
\newcommand{\mom}[2]{\bk{#1}_{\scriptscriptstyle \ns #2}}
\newcommand{\eqn}[1]{(\ref{eq.#1})}
\newcommand{\bra}[1]{\mbox{$\left\langle #1 \right|$}}
\newcommand{\ket}[1]{\mbox{$\left| #1 \right\rangle$}}
\newcommand{\braket}[2]{\mbox{$\left\langle #1 | #2\right\rangle$}}
\newcommand{\av}[1]{\mbox{$\left| #1 \right|$}}
\newcommand{\ans}{\mbox{$| a_n |^2$}}
\newcommand{\bk}[1]{\mbox{$\left\langle #1 \right\rangle$}}
\newcommand{\op}[1]{\mathsf{#1}}
\newcommand{\M}{{M}}
\newcommand{\W}{{\Delta\nu}}
\newcommand{\E}{{\Delta E}}
\newcommand{\p}{{\Delta p}}
\newcommand{\T}{{\Delta t}}
\newcommand{\X}{{\Delta x}}
\newcommand{\WX}{{\Delta\mu}}
\newcommand{\ie}{{\em i.e.}}
\newcommand{\eg}{{\em e.g.}}
\newcommand{\cf}{{\em cf.~}}
\newcommand{\ts}{\textsuperscript}
\newcommand{\sss}{\scriptscriptstyle}
\newcommand{\vara}{{\mbox{\sl a}}}
\newcommand{\half}{{1\ns\ns/\ns2}}
\newcommand{\propsep}{{\vspace{0em}\noindent}}
\newcommand{\proof}{{\propsep\em Proof: }}


\makeatletter \newcommand\prop[1]{\propsep
  \@startsection{paragraph}{4}{0pt}{\z@}{-1em}{\bf} {Proposition #1:}
  \phantomsection \renewcommand{\@currentlabel}{#1} \label{prop.#1}}
\makeatother

\title{The maximum distinctness of physical systems}

\begin{abstract}
The limited distinctness of physical systems is roughly expressed by
uncertainty relations. Here we show distinctness is a finite resource
we can exactly count to define basic physical quantities, limits to the
resolution of space and time, and informational foundations for
classical mechanics. Our analysis generalizes quantum speed limits: we
count the distinct (orthogonal) states that can occur in a finite
length of unitary change.  As in Nyquist's bound on distinct signal
values in classical waves, widths of superpositions bound the distinct
states per unit length---and basic conserved quantities are widths.
Maximally distinct unitary evolution is effectively discrete---and this
characterizes classical systems.

\end{abstract}

\author{Norman Margolus}

\affiliation{Massachusetts Institute of Technology, Cambridge MA 02139}

\maketitle

\section{Introduction}

We live in a quantum world in which distinctness is a finite resource
and counting distinct (orthogonal) states defines basic quantities such
as entropy and energy.

Entropy was the first of these quantities to be recognized as a count
of states (or conventionally as the $\log$ of a count).  This was
discovered by Boltzmann, who coarse-grained classical-mechanical
state-space in order to count states and apply statistics to thermal
systems~\cite{boltzmann2}.  Planck extended Boltzmann's statistics to
the interaction of light and matter, and found entropy matched
experiment if the grain-size had a particular value
$h$~\cite{planck,sackur-tetrode}. Entropy was an {\em absolute} count.
As the wave nature of momentum and energy became apparent
\mbox{\cite{einstein,deBroglie,schrodinger}}, limited distinctness was
explained as a wave property, like the minimum product of spatial and
spatial-frequency widths for classical wave\-packets
\cite{heisenberg,bohr}.  This has been formalized in uncertainty
relations \cite{kennard,folland}, but these only roughly define a count
of distinct quantum states in classical state space.

Here we introduce new bounds that tell us exactly how many
\textit{perfectly distinct} (orthogonal) \textit{quantum states} can
occur in a \textit{classical length} of physical change. They show that
counting states is as fundamental in the rest of physics as it is in
thermal systems. For example, energy counts how many distinct states
can occur per unit time, and momentum counts those allowed by~a unit
length of motion.  Classical mechanics approximates a maximally
distinct quantum evolution, making it an interplay of counts and
effectively discrete.  Finite distinctness limits the resolution of
measurements and of approximately classical spacetime. It links
mechanics and information, continuous and discrete, and classical and
quantum.


The prototype for new distinctness bounds is Nyquist's classical bound
on communication with waves~\cite{nyquist2}.  He showed that the
critical resource that limits the number of distinct signal values that
can be transmitted per unit time is \mbox{\em bandwidth}, the width of
the range of frequencies that can appear in the wave's Fourier
decomposition.  Intuitively, doubling bandwidth lets us double all
frequencies, making everything happen twice as fast---including
distinct values.  Energy is similarly the critical resource that limits
distinct change in quantum time evolution: doubling all energies would
make evolution twice as fast.

\begin{figure}[t]
    \includegraphics[width=2.3in]{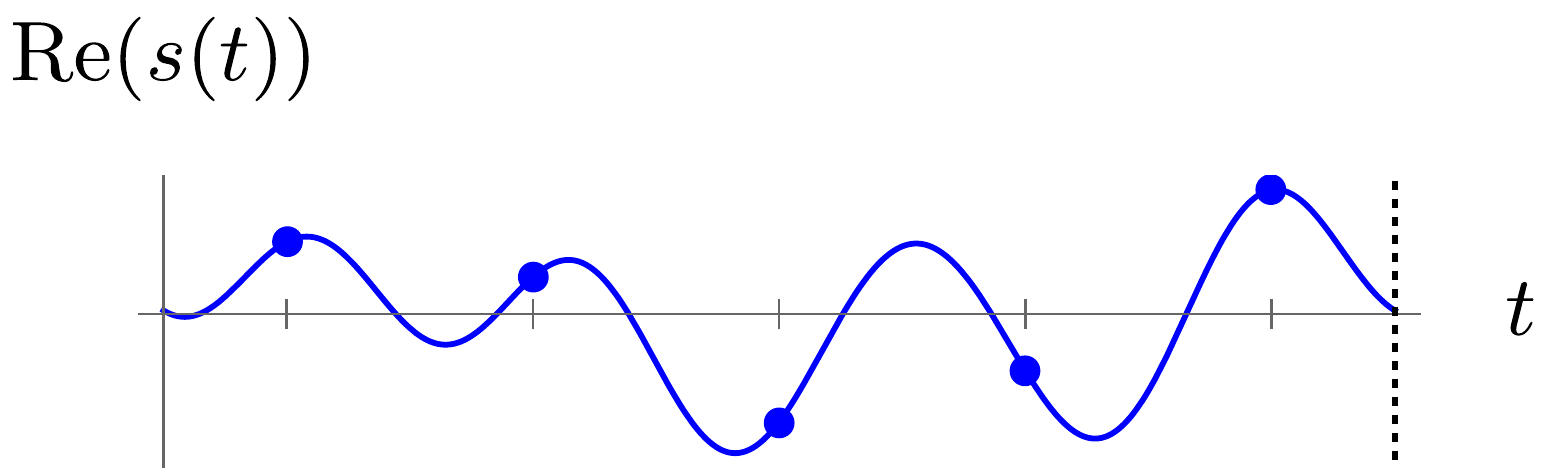}
    \caption{ {\it A wider frequency range allows more distinct
        values.}  We show one period of ${s(t)}$, a sum of five Fourier
      \mbox{components} with consecutive numbers of whole cycles per
      period.  Values chosen at five different times determine all
      coefficients in the sum, hence $s(t)$ at all times.  {\em $N$
        frequencies allow $N$ choices}.}
\label{fig.finite-bw}
\end{figure}

\begin{figure}[t]
    \includegraphics[width=2.3in]{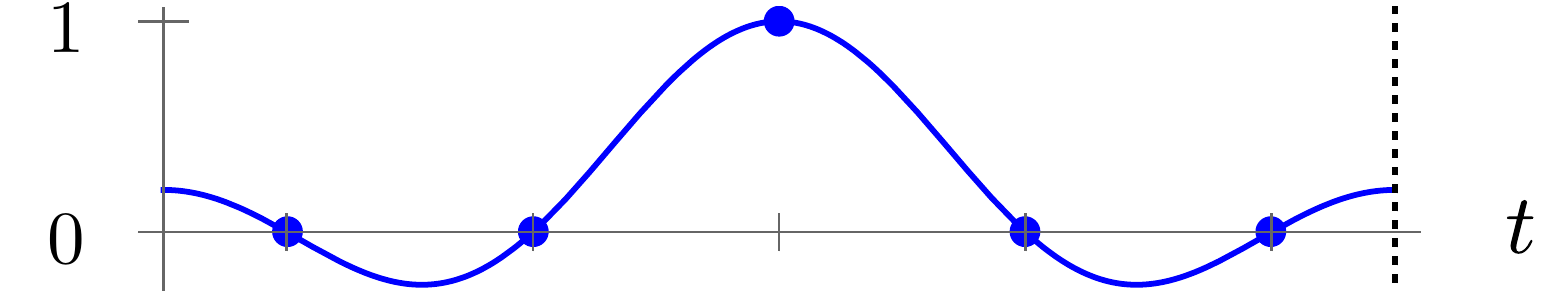}
    \caption{\textit{Distinct times define localized waves.}  With $N$
      distinct values at equally spaced times we can construct $N$
      waves,~each zero at all but one of the times.  These form a basis
      for \textit{any wave} with the same period and frequency range:
      in a sum, each lets us set one distinct value without affecting
      the rest.}\label{fig.packet}
\end{figure}

\begin{figure}[t]
    \includegraphics[width=2.2in]{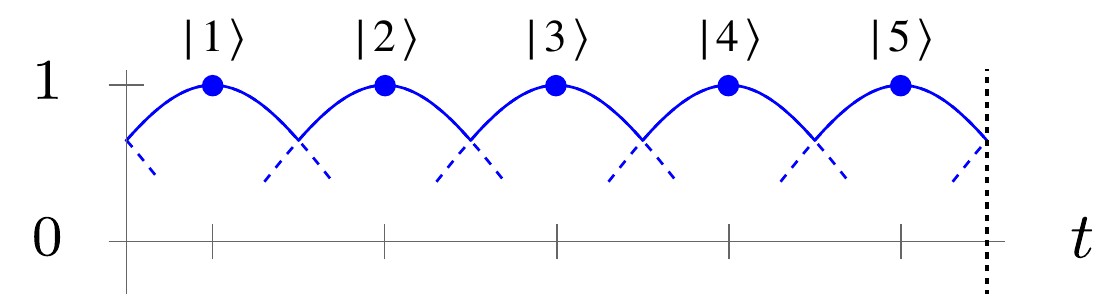}
    \caption{{\it Fastest evolution through distinct vectors is
        unitary}.  To define a vector evolution with fastest distinct
      change for a finite frequency range, multiply localized waves by
      distinct {\em orthonormal} vectors. The sum will be normalized at
      all times.}\label{fig.unitary}
\end{figure}

\begin{figure}[t]
    \includegraphics[width=.75\columnwidth]{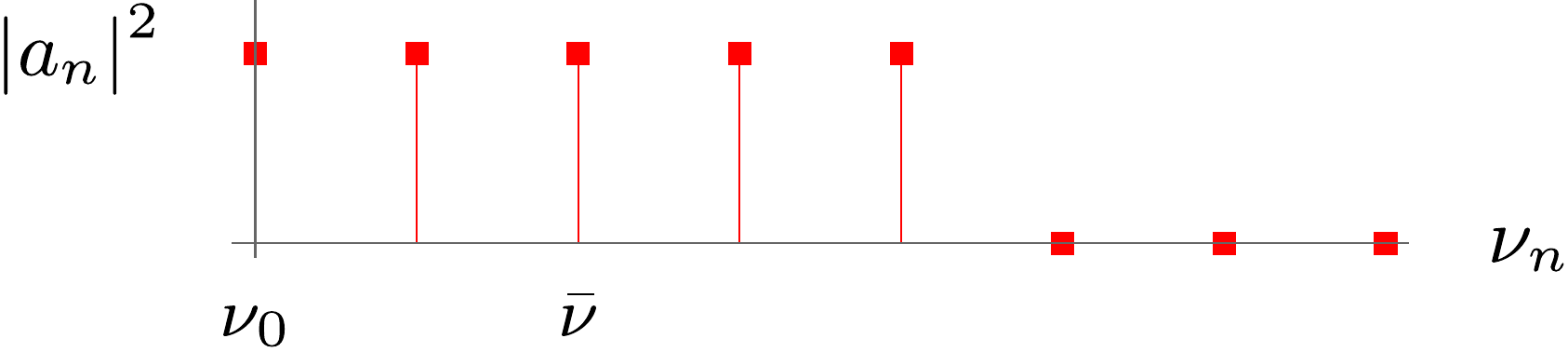}
    \caption{{\it Distinctness bounds are bandwidth bounds.}  The most
      distinct evolution a width $\W$ allows has the widest possible
      evenly-weighted range of frequencies. \textit{E.g.}, for
      $\W=\bar\nu-\nu_0=$
      $\textit{average}\ps-\ns\textit{ground-state},\ns\ns$ its value
      tells us $\bar\nu\ns,$ widest starts at $\nu_0.$
    }\label{fig.numin-numax}
\end{figure}

\paragraph{\bf To briefly review,} Nyquist's classical bandwidth bound
derives from the fact that a periodic wave has a discrete spectrum with
only a finite number of frequencies in a finite range.  For example,
consider a {\em complex-valued signal} $s(t)$ with period $T$.  Only
frequencies $n/T$ that cycle an integer number $n$ times per period
appear in its Fourier series: $s(t)\!=\!\sum c_n \, e^{2\pi i t \ps
  n\ns/T}$. If the sum has just~$N$ terms, all coefficients $c_n$ are
determined by choosing the value of the signal at $N$ times.  This
defines maximum distinctness: $N$ chosen signal values with $N$
frequencies (Figure~\ref{fig.finite-bw}).  Since allowed frequencies
are $1/T$ apart, any frequency range that includes $N$ frequencies has
width
\begin{equation}\label{eq.wt}
      {\nu_{\max} - \nu_{\min}} \ge \frac{N-1}{T}\;.
\end{equation}
Thus for a long signal its {\em bandwidth}, defined here as $\nu_{\max}
- \nu_{\min}\ps$, is the maximum average number of chosen values per
unit time, $N/T$~\cite{bw-note}.  Moreover, finitely spaced signal
values determine all coefficients in the Fourier sum, so the continuous
signal is completely determined by a discrete subset of its
values~\cite{interpolation}: a finite bandwidth wave in time (or
similarly in space) is {\em effectively discrete}.

With $N$ distinct values at equally spaced times we can construct $N$
{\em localized waves}, each zero at all but one of the times, where it
is one (Figure~\ref{fig.packet}).  These localized waves form a basis
for constructing \textit{any wave} with the same period and frequency
range. In a superposition, each lets us set one distinct value without
affecting the rest: we multiply the localized wave by the distinct
value.

If we instead multiply each localized wave by a \textit{distinct
  vector}, we define a \textit{vector evolution} with components that
are waves, all with the same period and frequency range
(Figure~\ref{fig.unitary}). If the distinct vectors are orthonormal,
the superposition remains normalized at all times: the fastest vector
evolution \eqn{wt} allows is \textit{unitary} (see
Appendix~\ref{appendix.interpol}).

\begin{figure}[t]
  \includegraphics[width=2.2in]{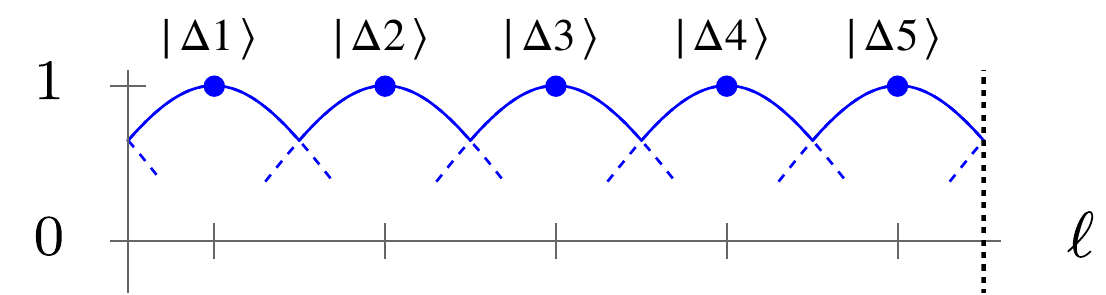}
  \caption{{\it Many unitary transformations are like time-evolution.}
    A parameter $\param$ plays the role of time in the transformation.
    Waves localized at distinct ``times'' are multiplied by
    distinctly changed $\ket{\Delta n}$. For shifts in space,
    $\param$ is amount of~shift.}\label{fig.unitary-l}
\end{figure}

\begin{figure}[t]
  \includegraphics[width=1.6in]{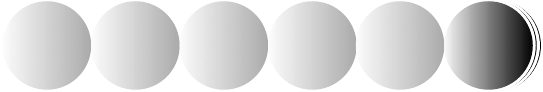}
  \caption{{\it 
      Some states are distinct due to motion.}  A localized system
    viewed in a uniformly moving frame traverses distinct states not
    seen in its rest frame.  Overall momentum---also due entirely to
    frame motion---bounds the extra distinctness.}
\label{fig.moving-p}
\end{figure}

\paragraph{\bf This paper extends these ideas.} Quantum time
evolution is unitary, with rate of change governed~by energy
eigenfrequencies $\nu_n=E_n/h$.  As in Figure~\ref{fig.unitary}, a
finite frequency range determines a maximum rate of distinct orthogonal
change.  Although the spectrum of energies available in a quantum
dynamics may be constrained, we can ignore that in constructing general
energy bounds, since it is the least constrained cases that allow the
most distinctness.  Thus \eqn{wt} applies also to periodic quantum
evolutions.  Rewritten in terms of energy and $\tau=T/N$, the average
time between distinct states, \eqn{wt} becomes
\begin{equation}\label{eq.de-t}
\frac{E_{\max}-E_{\min}}{h}\,\tau \ge \frac{N-1}{N}\;.
\end{equation}
As the number $N$ of distinct states increases, \eqn{de-t} becomes
\textit{independent of period and periodicity}.  Other definitions of
energy width $\E=h\ps\W$ give similar bounds
\begin{equation}\label{eq.dnu-t}
\W\,\tau \;\gtrsim \;1
\end{equation}
as long as the same intuition applies: doubling $\W$ lets us double all
frequencies, hence all rates.  In fact, all bounds \eqn{dnu-t} are
really bandwidth bounds: we show we always get the most distinctness in
time by using the \textit{widest range} of \textit{evenly weighted}
frequencies the definition and value of $\W$ allow, and this
\textit{finite bandwidth} is bounded by \eqn{de-t}.

We illustrate this in Figure~\ref{fig.numin-numax} for
$\W=\bar\nu-\nu_0\ps$: \mbox{average} minus ground state.  The {\em
  value} of $\W$ tells us $\bar\nu$; an evenly weighted range about
$\bar\nu$ is at most $2\W$ wide; and from~\eqn{de-t}
\begin{equation}\label{eq.etau}
\frac{2(E-E_0)}{h}\,\tau \ge \frac{N-1}{N}\;.
\end{equation}
This generalizes an achievable bound~\cite{max-speed} on how fast an
evolution with \textit{average energy} $E$ can transition between $N=2$
distinct states: a fastest evolution through $N$ distinct states is
periodic with even spacing $\tau$ \mbox{(\cf
  \cite{mandelstam,uffink85,hilg96,braunstein,yu,max-speed,luo2,g0,luo1,zych,ness,chau})}.
This is also a bound on the density of \textit{distinct moments of
  time} in an evolution, hence on the maximum resolution of time
\mbox{measurements}
\mbox{(\cf\cite{caves,yurke,vourdas,sanders,ou,kok,g1,durkin,g3,hall,zwierz2,caves2})}.
For the total special \mbox{relativistic} energy of a system, the
lowest energy $E_0=0$, so for $N$ large and $h=2$, \eqn{etau} is simply
$1/\tau\le E$.  This identifies total energy as the total rate of
distinct change possible in all dynamics at all scales.  This is also
the fastest rate of classical information change
(\cf\cite{brem-comm,bek-comm,deffner-comm,bek-ratio,ivanov,casini,longo}).

Other unitary transformations are like an evolution in time, but with
some other classical parameter $\ell$ giving the amount of
transformation (Figure~\ref{fig.unitary-l}).  For example, $\ell$ might
be the length of a shift in space, with \mbox{momentum} (spatial
frequency) playing the role of energy; or $\ell$ an angle of rotation,
with angular momentum the frequency.  We can relate distinctness
possible in any transformation to energy bounds by imagining we
transform the state at a constant rate, so it becomes a time evolution:
$\ell\propto t$.

The example of a constant-rate shift of classical space coordinates is
particularly relevant for special relativity, since it defines an
\textit{inertial frame} (Figure~\ref{fig.moving-p}).  Most of the
dynamics of a well-localized isolated system, including all
wavefunction spreading, can be described in its rest frame.  Its
overall motion and momentum can be described separately, as a pure
shift: a sum of plane waves that all move in the same direction at the
same speed.  Then the momentum analog of \eqn{etau}, with lowest
momentum $p_0=0$ and $N$ large, identifies average momentum $p$ as the
maximum number of distinct changes per unit shift in space, just as
relativistic energy $E$ is in time.  Thus if we model classical
mechanics as quantum evolution that is \mbox{\em maximally} \mbox{\em
  distinct} (rather than infinitely distinct), the classical
relativistic relationship between two frames,
\begin{equation}\label{eq.intro-rel}
  E\ps\T - p\ps\X = E_{\text{rest}}\ps\T_{\text{rest}}\;,
\end{equation}
relates counts: total distinct-changes between two events $\T$ apart,
minus those due to motion of length $\X$, equals those seen in the rest
frame.  Lagrangian action similarly counts rest-frame distinct changes,
so the principle of stationary action (over short times~\cite{gray})
becomes {\em maximum aging}~\cite{taylor}: maximum distinct-events in
rest frames.

Finally, a remarkable consequence of finite distinctness is
\textit{effective discreteness}: a maximally distinct evolution has
finite energy and momentum bandwidth, so the state for a discrete set
of times and positions defines the entire evolution.  This kind of
discreteness in no way precludes continuous
symmetry~(\cf\cite{kempf1,kempf2,kempf3,kempf4,tsang,marg-mech,emulation}), since
it has no fixed origin or orientation, but it does erase differences
between continuous and discrete evolution.  For classical dynamics
modeled as maximally distinct quantum evolution, the spacetime
discreteness scale is determined locally by the average energy density;
only at the Planck energy density does it become the Planck length.
Effective discreteness also lets classical lattice gases play the same
kind of role in the rest of mechanics they do in statistical mechanics.

\paragraph{\bf The plan for this paper} is to prove general bounds on
distinctness, then provide examples of their striking consequences,
particularly for the relationship of classical to quantum, and conclude
with a high-level view of their significance.  Appendices discuss
\ref{appendix.interpol}: interpolation, \ref{appendix.numerical}:
numerical \mbox{methods} used in the Figures and in supplementary
verification code~\cite{code}, \ref{appendix.portion}: bounds on part
of an evolution, \ref{appendix.large}: expected distinctness of large
evolutions if we include all related entanglement,
\ref{appendix.field-theory}: a classical lattice gas modeled as a
quantum field theory, and \ref{appendix.analysis}: effectively discrete
integration.  Some results were previewed in~\cite{unpublished}.

\section{Counting distinct states in time}

In quantum mechanics, orthogonal vectors represent \textit{distinct
  states}: states that can be distinguished from each other with
certainty.  A finite system with finite energy has only a finite number
of distinct states. Here we count how many distinct states {\em can
  occur in a finite time}.  This depends only on the wavefunction's
energy distribution.

\subsection{Only frequencies matter}

Consider a finite-sized isolated system in flat spacetime.  We can
express its time evolution as a superposition of discrete energy
(frequency) eigenstates:
\begin{equation}\label{eq.kett}
\ket{\psi(t)} = {\sum\nolimits_n a_n \,e^{- 2\pi i \nu_n
    t}\,\ket{E_n}}\;,
\end{equation}
with $\nu_n=E_n/h$.  For a normalized state the \ans{} add up to one
and play the role of probabilities for each $\nu_n$.  This lets us
define an average width $\W$ for the probability distribution---for
example, the standard deviation.

If the time evolution \eqn{kett} passes through a sequence of mutually
orthogonal states $\ket{\psi(t_k)}$ at times $t_k$, then
\begin{equation}\label{eq.orthogt}
\braket{\psi(t_m)}{\psi(t_k)}={\sum\nolimits_n \av{a_n}^2 \,e^{2\pi i
    \nu_n(t_m-t_k)} = \delta_{mk}}\;.
\end{equation}
Thus the frequencies $\nu_n$ are the \textit{only} characteristic of
the dynamical law that constrains orthogonal evolution.

Given a definition of average frequency width $\W$, we count the
\textit{maximum number} $N$ of distinct states that can occur for a
given value of $\W$ in a given time.  We do this by finding the
\textit{minimum value} of $\W$ for a given $N$.

\subsection{Defining frequency width}

We define {\em a well-behaved average frequency-width $\W$} to be a
non-negative function of a discrete set of frequencies and their
assigned probabilities, with Properties:
\begin{enumerate}\itemsep-2pt
  \item {\em Scales with Frequency}. It is multiplied by $\kappa$ if
    all frequencies are multiplied by $\kappa>0$.\label{p1}
  \item {\em Measures the Spread}. It is a function only of frequency
    differences, not absolute frequencies.\label{p2}
  \item {\em Weights Frequencies}. It does not change if
        probability is redistributed among equal frequencies.\label{p3}
  \item {\em Centered}. It does not decrease if probability
        mass is moved farther from some central frequency
        $\alpha$.\label{p4}
\end{enumerate}
\vspace{-.15em}
\noindent  
We call the width {\em natural} if it also has the Property:
\vspace{-.15em}
\begin{enumerate}\addtocounter{enumi}{4}
   \item {\em Natural}. It is of the same order of magnitude as
     bandwidth for a uniform probability distribution.\label{p5}
\end{enumerate}
\vspace{-.15em}
\noindent For a width to have units of frequency, it should scale with
frequency.  For identical frequencies associated with different states,
only their weights affect the distribution.  A rectangular distribution
has a natural width.

For example, if $\alpha$ is the largest, smallest, or average
frequency, the \textit{generalized deviation}
\begin{equation}\label{eq.dnu} 
    \mom{\nu-\alpha}{\M} \equiv {\Bigl(\sum\nolimits_n \av{a_n}^2 \,
      \av{\nu_n-\alpha}^\M\Bigr)^{\frac{1}{\M}}}\;,
\end{equation} 
\noindent $\M$\ts{th} root of $\M$\ts{th} moment of absolute deviation
from $\alpha$, has Properties \ref{p1}--\ref{p4} for $\M>0$.  Twice the
generalized deviation is a natural width with Property~\ref{p5} for
$\M\gtrsim 2/3$.

\subsection{All frequencies are allowed}

We count the number of distinct states possible in an isolated unitary
evolution with given frequency width and evolution length.
\textit{Disallowing} some frequencies cannot increase this count, since
we are free to just not use them.  \textit{Degeneracy} cannot help,
since moving probability \ans{} between states with identical
frequencies does not affect orthogonality \eqn{orthogt} or widths
(Property~\ref{p3}).  Thus we can establish bounds on {\em all systems}
by considering only ones with an unconstrained non-degenerate spectrum.

Now, an isolated unitary evolution is either periodic up to an overall
phase (which does not affect distinctness), or arbitrarily close to
periodic \cite{qm-recurrence,unified-recurrence}.  To count distinct
states possible in a time evolution with period $T$, the frequencies
that can appear in wavefunction \eqn{kett} are
\begin{equation}\label{eq.nun}
\nu_n=\nu_0 + n/T\;,
\end{equation}
with $n$ a non-negative integer.  Time evolution always has a lowest
frequency $\nu_0$ \cite{wightman}, and these are all the frequencies
with period $T$ and a non-degenerate spectrum.

To count the distinct states possible in a \textit{finite portion of
  any evolution}, the maximum period $T$ is unbounded so all
$\nu\ge\nu_0$ and all \textit{first recurrence times} are allowed.


\subsection{Bounds on an entire evolution}

The maximum rate of distinct change for a periodic or unbounded time
evolution with a finite {energy-frequency width} $\W$ is achieved by
using the \textit{widest range of evenly weighted energies} the value
of $\W$ allows.  The evolution is also maximally distinct for
\textit{all other widths} with values that correspond to this energy
distribution.  We evaluate bounds analytically below and verify
numerically in \cite{code}.


\vspace{.2em}

\prop1 Given an isolated periodic quantum evolution that traverses $N$
distinct states in period $T$, there is a smallest possible range of
energy eigenfrequencies $\nu_n=E_n/h$ of the state represented in the
energy basis: a {\em minimum bandwidth}
$\nu_{\max}-\nu_{\min}\ge(N-1)/T$.

\proof At least $N$ distinct $\ket{E_n}$ must appear in the
superposition to add up to $N$ distinct states $\ket{\psi(t_k)}$ at
different times $t_k$.  Since the frequencies \eqn{nun} compatible with
the periodicity differ by at least $1/T$, the minimum possible
bandwidth to have $N$ of them is $(N-1)/T$.~\hfill$\square$

\prop2 Minimum bandwidth is only possible~if the $N$ distinct states in
period $T$ are evenly spaced.

\proof A minimum bandwidth superposition \eqn{kett} has just $N$
non-zero $a_n$, so is only possible if there are at most $N$ linearly
independent constraints on the \ans{}.  From \eqn{orthogt}, the times
$\delta t \le T/2$ separating {\em any two} of the $N$ distinct states
constrain the \ans{}. ($T-\delta t$ and $\delta t$ are equivalent).
For $0<\delta t<T/2$, each different $\delta t$ gives two linearly
independent constraints: real and imaginary parts of \eqn{orthogt}.
$\delta t= 0$ or $T/2$ gives only one constraint.  Even spacing, $T/N$
apart, gives the fewest different separations $\delta t$ and exactly
$N$ constraints.  Uneven spacing gives more.~\hfill$\square$

\prop3 A minimum-bandwidth superposition that traverses $N$ distinct
states in period $T$ contains $N$ frequencies $\nu_n$, each with
probability weight $|a_n|^2=1/N$.

\proof With equal spacing $\tau= T/N$ between distinct states, let
$t_k=k\ps\tau$ for $0\le k\le N-1$.  From \eqn{nun} and \eqn{orthogt},
\begin{equation}\label{eq.nsum}
\braket{\psi(t_{k})}{\psi(t_0)}= {e^{2\pi i \nu_0 t_k}
  \sum\nolimits_{n=0}^\infty \av{a_n}^2 \,e^{2\pi i n k/N}.}
\end{equation}
Since there are only $N$ different phase factors in the sum, we can
pick any $N$ consecutive values of $n$ and provide one non-zero
coefficient \ans{} for each phase factor.  Then, since
$\braket{\psi(t_{k})}{\psi(t_0)}\ns=\ns\delta_{k 0}\ns=\ns e^{2\pi i
  \nu_0 t_k}\delta_{k 0}$, the non-zero \ans{} are just the discrete
Fourier transform of a Kronecker delta impulse, and so they all equal
$1/N$.~\hfill$\square$

\prop4 With $N$ distinct states spread evenly in time, any
\textit{average width} $\W$ centered on $\alpha$ is minimized by a
minimum-bandwidth superposition centered on $\alpha$.

\proof As shown above, with an even spread in time there are just $N$
different phase factors in \eqn{nsum}, and \eqn{orthogt} can only be
satisfied if they all have equal probability weight~$1/N$.  Suppose we
put this weight on $N$ consecutive $\nu_n$ centered on $\alpha$.  Then,
from Property~\ref{p4} of $\W$, no rearrangement of probability weights
that keeps the same total probability $1/N$ on each different phase
factor can decrease $\W$.~\hfill$\square$

\prop5 With equal spacing $\tau=T/N$ between consecutive distinct
states and a $\W$ centered on $\alpha$,
\begin{equation}\label{eq.ntf}
\W\,\tau \ge f\,,
\end{equation}
where $f$ is independent of $T$.  We get equality for any minimum
bandwidth superposition centered on $\alpha$.

\proof From \eqn{nun} and Properties \ref{p1} and \ref{p2}, $\W\propto
1/T$ so $\W\,\tau$ is independent of $T$ and so is its
minimum. Given~$\tau$, minimizing $\W$ (Proposition~\ref{prop.4})
minimizes $\W\,\tau$.~\hfill$\square$

\begin{figure}[t]
  \includegraphics[height=.4\columnwidth]{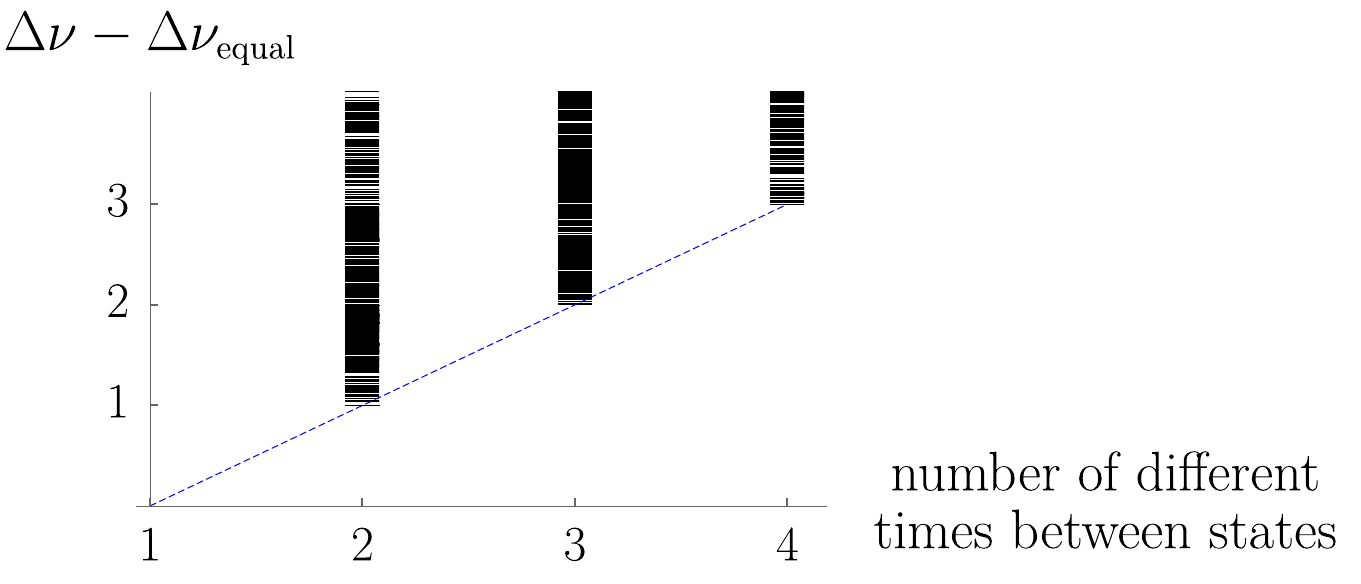}
    \caption{{\it Equal time between distinct states allows
        smallest~$\Delta \nu$}. Each mark shows a numerical minimum of
      $\Delta\nu=2(E-E_0)/h$ for distinct states placed randomly in
      period $T\ns = 1$ with $\le 4$ {\em different times} between
      them. $\Delta\nu_{\text{equal}}$ is $\min$ with \textit{equal
        times}.}
\label{fig.eq-is-min}
\end{figure}

In fact, \eqn{ntf} still holds for $\tau=T/N$ if distinct states are
\textit{unevenly spaced}. This adds constraints \eqn{orthogt} that, as
in Proposition~\ref{prop.2}, increase the minimum.  Even tiny
departures from evenness increase the minimum discretely: as we let
times between adjacent states converge, we get even-spacing constraints
{\em plus} additional ones~\cite{uniformity-note}.  This is illustrated
for $\W=2(E-E_0)/h$ in Figure~\ref{fig.eq-is-min}, and for a wide
variety of other $\W$ in \cite{code}.  Sets of distinct states are
placed randomly in period $T=1$, with up to 4 different times between
adjacent states in each set. Each mark shows the minimum $\W$ possible
for one set---numerical methods are discussed in
Appendix~\ref{appendix.numerical}.  Smallest~minima (dashed line) are
achieved as the times \textit{approach} equality.

\prop6 For any natural measure $\W$ of average frequency-width, $f$ has
order of magnitude one.

\proof Since $\W$ is minimized by a uniform minimum-bandwidth
distribution, from Property~\ref{p5} its minimum has magnitude $\sim
(N-1)/T$ so $f \sim (N-1)/N \sim 1$.~\hfill$\square$

\prop7 $\ns\!\!$Distinctness is maximized by the \textit{widest range}
of evenly-weighted frequencies a given $\W$ allows.

\proof The maximum $N$ allowed by a given $\W$ and $T$ minimizes
$\W\,T/N=\W\,\tau$.  As shown above, an evenly-weighted frequency range
achieves the minimum.~\hfill$\square$

Thus any evolution with an evenly-weighted frequency range starting at
$\nu_0$ is maximally distinct for its average energy above the lowest
possible (Figure~\ref{fig.numin-numax}) and for \textit{all widths
  compatible with that average energy}
(\cf\cite{tight-note,toffoli-levitin,chau-forbidden}).

\subsection{Examples of periodic-evolution bounds}

\begin{figure}[t]
  \includegraphics[height=.4\columnwidth]{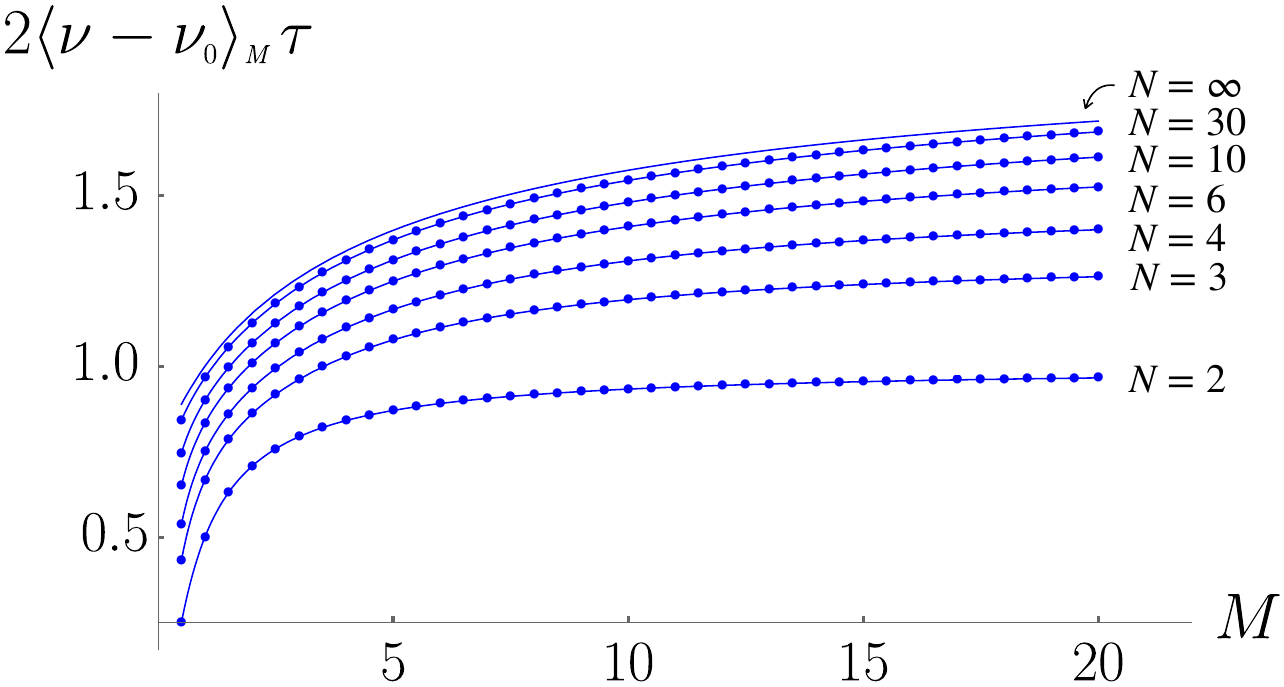}
  \caption{{\it Verifying generalized deviation bounds.}  Periodic
    evolution bounds $f_{\nu_0}(\M,N)$ (solid) exactly match numerical
    minima (dots) for an entire evolution or a portion
    \mbox{containing} $N$ distinct states $\tau$ apart.  $f_{\bar\nu}$
    match portion-minima if $M\ge2$.}
\label{fig.theory-vs-data}
\end{figure}

Let $\W=2\ps\ps\mom{\nu-\alpha}{\M}$, twice the average
deviation~\eqn{dnu} of $\nu$ from $\alpha$ for moment $\M>0$.  For an
evolution with period $T$, $N$ distinct states, and $\tau=T/N$,
\eqn{ntf} becomes
\begin{equation}\label{eq.fn}
2\,\mom{\nu-\alpha}{\M}\,\tau \;\ge\; f_\alpha(\M,N)\;,
\end{equation}
where $f$ is independent of $T$ and depends only on $N$ and the
parameters $\alpha$ and $\M$ that define $\W$. We get equality for a
range of $N$ evenly weighted frequencies \eqn{nun} centered on
$\alpha$.  For example, for deviations from $\alpha=\nu_0$ the range
starts at $\nu_0$, so from \eqn{dnu} the minimum of
$2\ps\ps\mom{\nu-\nu_0}{\M}\ps\tau$ is
\begin{equation}\label{eq.fnu0}
f_{\nu_0}(\M,N) = 2\, N^{-(1+\frac{1}{\M})}
\,\left(\ps{\sum\nolimits_{n=0}^{N-1} \, n^\M}\right)^{\frac{1}{\M}}\;,
\end{equation}
varying from $1/2$ to 2 for $\M\ns\ge\ns1$, with
$f_{\nu_0}(\M,2)=2^{-\frac{1}{\M}}$.  See
Figure~\ref{fig.theory-vs-data} (solid).  We get the same bounds if
some other $\alpha=\nu_n$ is the lowest with weight in the~wavefunction
or if $\alpha$ is a {\em highest} $\nu_n$ (deviations are absolute
values).

Another interesting case is a width about the mean frequency:
$\alpha=\bar\nu$.  An equally weighted range is always centered on its
mean, so the minimum of $2\ps\ps\mom{\nu-\bar\nu}{\M}\ps\tau$ is
\begin{equation}\label{eq.fnubar}
f_{\bar\nu}(\M,N) = 2\, N^{-(1+\frac{1}{\M})}
\left({{\sum\nolimits_{n=0}^{N-1}
    \av{n-\frac{N-1}{2}}^\M}}\right)^{\!\frac{1}{\M}}\!,
\end{equation}
varying from $4/9$ to 1 for $\M\ge 1$, with $f_{\bar\nu}(\M,2)=1/2$~for
all $\M$. For large $N$, convergence is rapid
(Figure~\ref{fig.theory-vs-data}) and
\begin{equation}\label{eq.finf} f_{\bar\nu}(\M,\infty) = 
\frac{1}{2}\, f_{\nu_0}(\M,\infty) =
\left(\frac{1}{1+\M}\right)^{\frac{1}{\M}},
\end{equation}
varying from $1/e$ to 1 for $\M>0$, independent of period or
periodicity.  For large $M$, $\mom{\nu-\nu_0}{\infty}$ is bandwidth
and
\begin{equation}\label{eq.fband} f_{\bar\nu}(\infty,N) = 
\frac{1}{2}\, f_{\nu_0}(\infty,N) = \frac{N-1}{N}\;,
\end{equation}
which varies from $1/2$ to 1 for $N\ge2$.

\newcommand{\prob}{{\text{prob}}}

\begin{figure}[t]
    \includegraphics[height=.4\columnwidth]{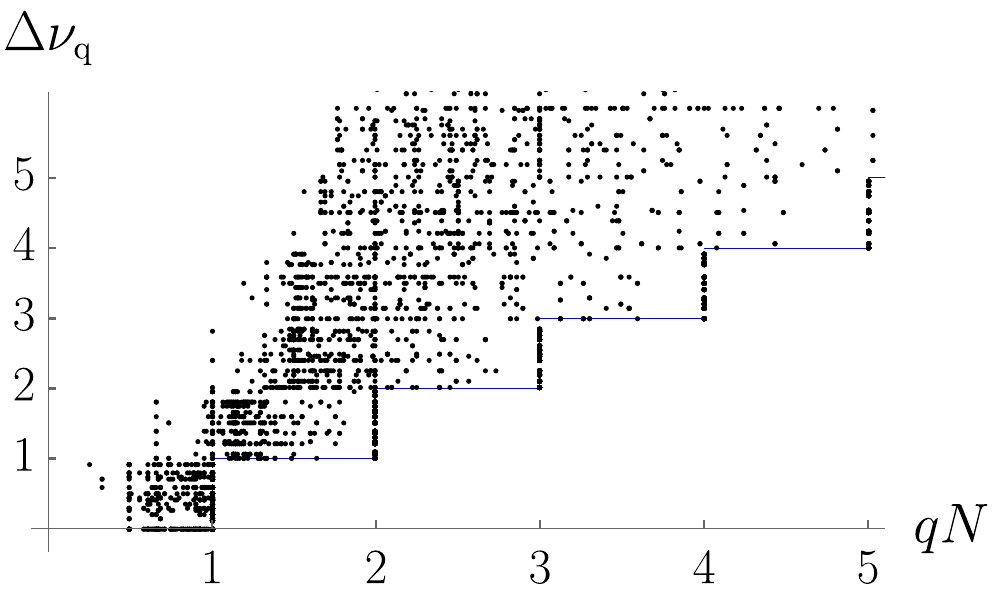}
    \caption{{\it Bandwidth required for total probability $q$.}  Each
      dot shows the width $\W_q$ of a smallest frequency range with
      total probability $q$, given $N\!\le\!  10$ distinct states
      placed randomly in period $T\! = \! 1$. The blue bound (equal
      spacing) is $N\ps f_\prob(q, N)$.}
\label{fig.stair}
\end{figure}

We can bound other widths similarly. For example, let
$\Delta\nu_q=\nu_{\max}-\nu_{\min}$ for a range of frequencies with
total probability $q$ (sum of weights is $q$) \cite{uffink85,hilg96}.
A {\em smallest range} with weight $q$ starts at an $\alpha=\nu_n$, and
\eqn{ntf} gives
\begin{equation}\label{eq.uffink}
\Delta\nu_q \,\tau \ge f_\prob(q,N)
\end{equation}
for some $f_\prob(q,N)$ independent of $T$.  Let $T=N$.  For equality
in \eqn{uffink}, $\Delta \nu_q\,\tau=\Delta \nu_q$ must encompass
$\lceil q N \rceil$ of $N$ evenly weighted frequencies $1/N$ apart, so
the bandwidth
\begin{equation}\label{eq.ceil-q}
f_\prob(q,N) = \frac{\lceil q N \rceil -1}{N}\;.
\end{equation}
In Figure~\ref{fig.stair} this bound (blue line) is tested
numerically. For $q\le 1/2$, $\Delta\nu_q$ is not a {\em natural
  width}, since then it can be zero for a discrete uniform
distribution; for $q>1/2$, $f$ ranges from 1/3 to 1.  Asymptotically,
$f_\prob(q,\infty)=q$.

\subsection{Bounds on a portion of an evolution}

What value of a width $\W$ is needed to have $N$ distinct states in a
\textit{given length} of time evolution? In general, the answer is that
$\W$ is smallest when the states are equally spaced, $\tau$ apart, and
the entire evolution has period $N\tau$: it obeys a periodic evolution
bound
(\cf\cite{mandelstam,uffink85,hilg96,braunstein,yu,max-speed,luo2,g0,luo1,zych,ness,chau}).
We can verify numerically that this is \textit{usualy} true for
deviation widths (Figure~\ref{fig.theory-vs-data}): it holds exactly in
all cases except for some deviations from the mean with $M<2$ and $N$
even.  Exceptions are studied in Appendix~\ref{appendix.portion}, but
the signature that a width is exceptional is that \textit{its periodic
  bound~can decrease with $N$}: this is incompatible with it applying
to all continuations of a portion.  For a time-independent Hamiltonian
$\op{H}$, periodic bound \eqn{fn} can be written
\begin{equation}\label{eq.fnh}
{2\, \bk{\av{\op{H}-E_\alpha}^\M}^{\frac{1}{\M}}\,\tau} \;\ge\;
h\,f_{\alpha}(\M,N)\;
\end{equation}
and this \textit{also governs portions} for all $f_{\alpha}$ monotonic
in~$N$.  The exceptional portion bounds can be smaller than $h\ps
f_{\alpha}$, but \textit{all bounds are order of magnitude} $h$ for
$\M\gtrsim 2/3$~\cite{code}.

\begin{figure}[t]
    \includegraphics[height=.4\columnwidth]{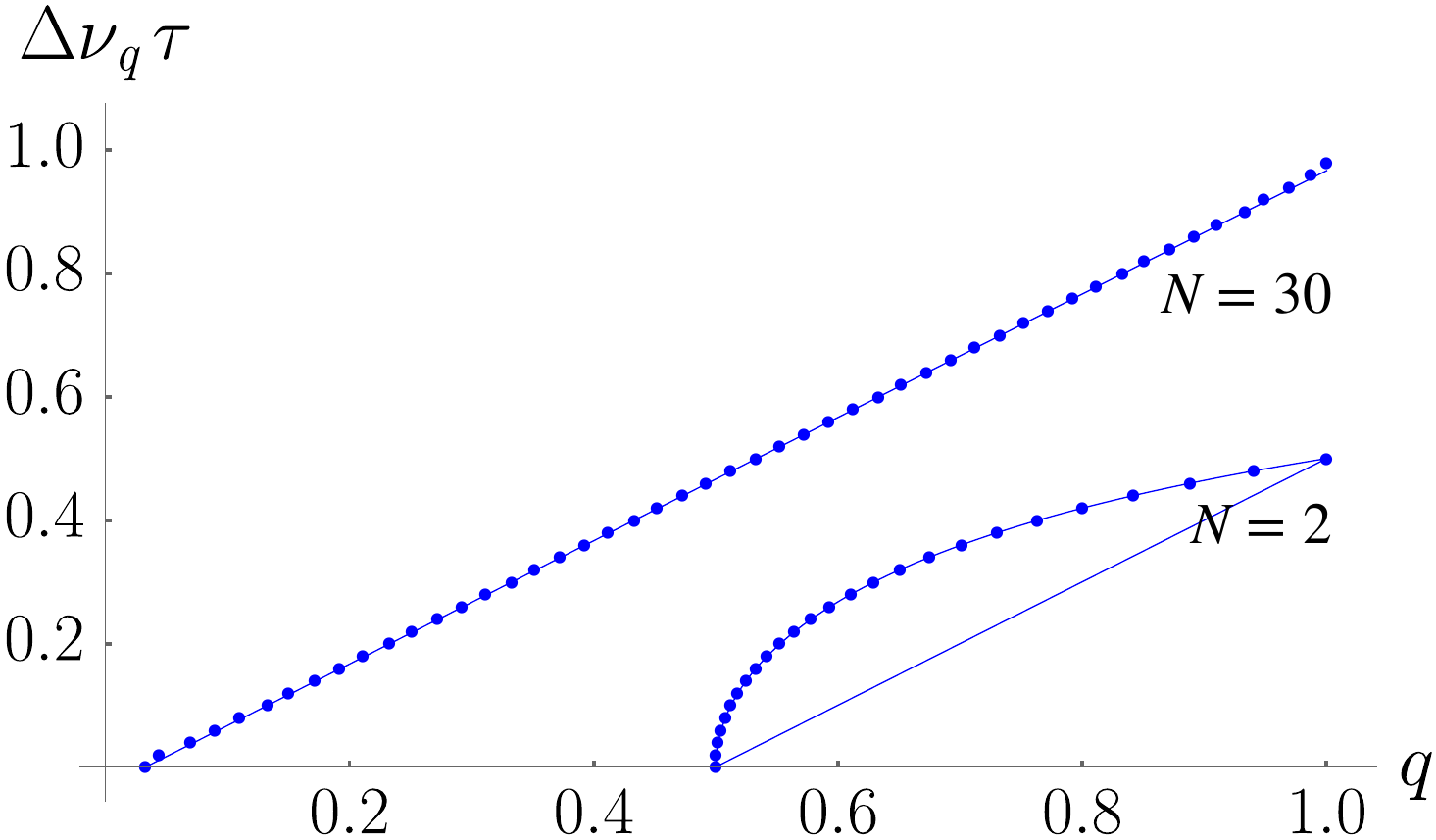}
\caption{{\it Minimum of $\Delta\nu_q\,\tau$ for a portion.}  For an
  evolution~in which a portion has $N$ distinct states $\tau$ apart,
  minimum bandwidth $\Delta\nu_q$ is needed for total probability $q$.
  Dots are numeric minima, curve an exact bound, straight lines
  $\Delta\nu_q\tau=q-1/N$.}
\label{fig.qmax}
\end{figure}

Our other example, the bandwidth $\Delta\nu_q$ of an energy range with
total probability $q$, is exceptional for $q<1$.  Periodic-evolution
bound \eqn{ceil-q} can decrease with $N$, so cannot apply exactly to
portions.  Figure~\ref{fig.qmax} compares~the best {\em linear bound}
implied by \eqn{ceil-q}, $\Delta\nu_q\tau\ge q-1/N$, to exact portion
bounds (dots) determined numerically.  The curved line is an achievable
portion-bound for $N=2$~\cite{uffink85}:
\begin{equation}\label{eq.arccos}
\Delta\nu_q\,\tau\ge \pi^{-1} \arccos (1/q\,-\,1)\quad\mbox{for}\quad
1/2\le q\le 1\;.
\end{equation}

\section{Other distinct transformations}

Energy bounds tell us how many distinct quantum states can occur in a
classical length of time evolution.  The same bounds apply to other
state transformations with a classical length-parameter, such as
spatial shifts and rotations: if we imagine the transformation
occurring at a constant rate, it becomes a time evolution.  Related
bounds on rates of change of observables and of arbitrary physical
processes are discussed in \cite{operator-flows,process-bounds}.

\subsection{Momentum bounds spatial shifts}

The analysis of distinctness under shifts of a system in space is a one
dimensional problem, identical in form to the problem already discussed
of evolution in time.  Consider, for example, a scalar particle
confined to a finite volume by periodic boundary conditions.  The
wavefunction is a sum of products of momentum eigenstates


\newcommand{\s}{{r}}

\begin{equation}\label{eq.ketxyz}
\ket{\psi} = \sum_{n_x n_y n_z} c_{n_x n_y n_z}
\,\ket{p_{n_x}}\ket{p_{n_y}}\ket{p_{n_z}}\;.
\end{equation}
If $\ket{p_{n_x}}$ is periodic in the $x$-direction with period $L_x$,
possible spatial frequencies are $\mu_{n_x}=p_{n_x}/h=n_x/L_x$, with
$n_x$ an integer. If the system is shifted a distance $\s$ in the $+x$
direction, the wave function becomes
\begin{equation}\label{eq.shifted}
\ket{\psi(\s)} = \!\!\!\sum_{n_x n_y n_z} \!\! c_{n_x n_y n_z} \,
e^{-\frac{2 \pi i}{h}\, p_{n_{\ns x}}
  \s}\,\ket{p_{n_x}}\ket{p_{n_y}}\ket{p_{n_z}}\; ,
\end{equation}
which follows from the form of the one-dimensional eigenstate
$\braket{x\ps}{\ps p_{n_x}}$.  Now if shifts of $\s_k$ and $\s_m$ give
orthogonal states for $m\ne k$, and $\av{a_{n_x}}^2 = \sum_{n_y n_z}
\av{c_{n_x n_y n_z}}^2$, we get
\begin{equation}\label{eq.orthogs}
\braket{\psi(\s_m)}{\psi(\s_k)}=\sum_{n_x} \av{a_{n_x}}^2 \,e^{2\pi i
  \ps\mu_{n_{\ns x}}(\s_m-\s_k)} = \delta_{mk}\;.
\end{equation}
This is identical in form to \eqn{orthogt} and so yields the same
one-dimensional minimization problem as before---with $\s$ and $\mu$
playing the roles of $t$ and $\nu$.

More generally, the total momentum operator $\vec{\op{p}}$ for any
isolated quantum system is defined to be the generator of spatial
shifts~\cite{generator}: the unitary operator $e^{-(2\pi i/h)\,
  \vec{\op{p}}\cdot\vec{\s}}$ shifts a wavefunction a fixed distance
$\vec{\s}$ in space with no other changes, as in \eqn{shifted}.  This
definition of $\vec{\op{p}}$ makes its average value a conserved
quantity in systems where a shift commutes with time evolution.  The
Hamiltonian operator $\op{H}$ similarly generates change in time: the
unitary operator $e^{-(2\pi i/h)\, \op{H}t}$ evolves the wavefunction
by an amount $t$ in time.  We can simply substitute one Hermitian
generator for another, and conclude momentum and energy bounds are
formally the same: if $\vec{\s}=\s\,\hat{x}$ then $\op{H}\to \op{p_x}$
and $t\to \s$.

We can gain some insight into the relationship between space and time
bounds by formally adding time to the shift evolution.  Imagine turning
off the actual dynamics and replacing it with $\op{H}=\op{p_x} v$, with
$v$ a constant speed.  This dynamics shifts the state a distance
$\lambda= v \tau$ in time $\tau$, since $\op{H} \tau=\op{p_x} \lambda$.
Periodicity in time becomes periodicity in space, distinct states in
time become distinct in space.  If $\tau$ is the average time between
distinct shifts, $\lambda=v \tau$ is the average distance.  Thus, for
example, substituting $\op{p_x} \lambda$ for $\op{H} \tau$ in \eqn{fnh}
gives bounds on $\Delta p_{\op{x}}\, \lambda$ of
\begin{equation}\label{eq.fnp}
2\, \bk{\av{\op{p_x}-p_{\op{x}\,\alpha}}^\M}^{\frac{1}{\M}}\,\lambda
\;\ge\; h\,f_\alpha(\M,N)\;.
\end{equation}
The bounds \eqn{fnp} have the same dimensionless $f_\alpha$ as the time
bounds---with $\alpha$ defined relative to the momentum distribution.
They also have the same applicability to a portion of {\em shift}
evolution, with the same exceptions for $\M<2$ about $\bar\mu$ given by
\eqn{tan}.  If there is no lowest or highest frequency for a spatial
superposition, bounds about mean $p_{\op{x}\,\bar{\mu}}=\bk{\op{p_x}}$
apply but not ones about min or max. Other energy bounds, such as
\eqn{uffink}, similarly apply.

Bounds \eqn{fnp} include the $N\gg1$ bandwidth bound that counts
distinct states in classical phase space~\cite{peliti,sackur-tetrode}
and Yu's $N\ns\ns= 2$ bound~\cite{yu},
$\langle\ps{(\op{p_x}-\bk{\op{p_x}})^2}\ps\rangle^{\half}\,\lambda
\,\ge\, h/4$. Luo's $N\ns\ns= 2$ bound \cite{luo2},
$\langle\ps\av{\op{p_x}}\ps\rangle\,\lambda \,\ge\, h/4\vara$ with
$\vara=1.1382\ldots\,$, is an exceptional bound about mean zero, given
by \eqn{tan}.

\subsection{Angular momentum bounds rotations}

\newcommand\J{\Delta J_{\ns\op{z}}}

There is no traditional uncertainty relation of the form
$\Delta\theta\,\J\gtrsim h$ between angle and angular
momentum~\cite{carruthers}.  This would require, for $\J \ll h/2\pi$,
that $\Delta\theta \gg 2\pi$, but a width of observable angles $\gg
2\pi$ has little physical meaning.  Periodic distinctness bounds {\em
  are} of that form, but avoid this obstacle because transformation
length is classical.  They bound a state-space area per distinct state
(\eg, $\E\, T\ns/N \gtrsim h$), so for given period, bound the width
needed for $N$ distinct states (\eg, $\E \gtrsim h\ps N/\ps T$).


As we did with shifts, here we formally make spatial rotation a special
case of time evolution by making the evolution length depend on time.
Now, much like $\vec{\op{p}}$, the total angular momentum operator
$\vec{\op{J}}$ is defined to be the generator of
rotations~\cite{generator}: $e^{-(2\pi i/h)\,
  \vec{\op{J}}\cdot\vec{\phi}}$ rotates the wavefunction by an angle
$\phi$ about an axis $\hat{\phi}=\vec{\phi}/\phi$.  This becomes an
evolution in time if $\phi=\omega\ps t$, where $\omega=\theta/\tau$ is
the ratio of average separation in angle and time between distinct
states. To express this as an evolution $e^{-(2\pi i/h)\, \op{H}t}$ we
let $\op{H}t=\vec{\op{J}}\cdot\vec{\phi}\,,$ giving $\op{H}=
\vec{\op{J}}\cdot\hat\phi\,\omega$.  Letting
$\op{J_z}=\vec{\op{J}}\cdot\hat\phi$, the component of $\vec{\op{J}}$
along the rotation axis, this becomes
$\op{H}\ps\tau=\op{J_z}\ps\theta$.  Making this substitution we can,
for example, rewrite \eqn{fnh} in terms of $\J\,\theta$ as
\begin{equation}\label{eq.fntheta}
2\, \bk{\av{\op{J_z}-J_{\ns\op{z}\ps\ps\alpha}}^\M}^{\frac{1}{\M}}\,\theta
\;\ge\; h\,f_\alpha(\M,N)\;.
\end{equation}
All state-rotations have period $2\pi$, up to an overall sign which
does not affect distinctness. First recurrence can be shorter due to
rotational symmetry; portion bounds are achievable only if $N \theta$
is an integer fraction of $2\pi$.

For a full $2\pi$ rotation with $N$ distinct states and $\J$ the
bandwidth, $\theta=2\pi/N$ and \eqn{fntheta} looks like \eqn{wt}:
\begin{equation}\label{eq.dj-theta}
\frac{J_{\ns\op{z}\ps\ps{\max}}-J_{\ns\op{z}\ps\ps{\min}}}{h}
\ge \frac{N-1}{2\pi}\;.
\end{equation}
Since eigenvalues of $\op{J_z}$ are $h/2\pi$ apart for both bosonic and
fermionic systems, \eqn{dj-theta} is achieved by any evenly weighted
superposition with $N$ consecutive eigenvalues.  Other bounds
\eqn{fntheta} are achieved by the same states.

\subsection{Distinctness limits measurement resolution}

Finite distinctness limits resolution in measurements.  For example,
suppose we want to measure the time between two states of a unitary
evolution with average energy $E_0+\Delta E$.  For maximum resolution
the evolution should have as many distinct states as possible to give
minimum spacing $\tau\approx h/2\ps\Delta E$ between distinguishable
moments.  Distinct states then form a discrete basis for the evolution,
as in Figure~\ref{fig.unitary}, so the time between any two states is
only defined up to a resolution $\tau$.


Since light is often used to probe systems of \mbox{interest}, maximum
density of distinct states in a periodic optical evolution is a key
constraint on measurements.  This is given by energy bounds.
Historically, fundamental \mbox{limits} in interferometry stem from a
photon number bound on phase
resolution~\mbox{\cite{caves,yurke,vourdas,sanders,ou,kok,g1,durkin,g3,hall,zwierz2,caves2}}.
\textit{For light of frequency} $\nu$, \mbox{average} number of photons
$\bar{n}$ determines $\Delta E= \bar{n} h\nu$, and if $N$ distinct
states are $\tau$ apart in period $T$, the angle between distinct
phases $\theta=2\pi\tau/T$.  Since $T\le 1/\nu$ and, from~\eqn{etau},
$\,\Delta E\,\tau\ge \frac{h}{2}\ps\frac{N-1}{N}$, the bound
$\theta\,\bar{n} \ns\ge\ns \pi\ps\ps\frac{N-1}{N}$ is achievable.
Since also $\theta\ns=\ns 2\pi/N$, this is equivalent to
$N\ns\ns\le\ns\ns 2\ps\bar{n}+1$. \textit{In general}, though, only
energy bounds distinctness in time.

For an ideal interferometer, maximally distinct evolution allows the
best resolution in time differences~\cite{durkin,noon-note}.  With
monochromatic light and beam splitters, overall evolutions are
isomorphic to 3D rotations~\cite{yurke,vourdas,sanders} and~so the
bounds \eqn{fntheta} apply.  In practice, highest resolution~is
currently achieved using {\em squeezed states} and evolutions that are
not maximally distinct~\cite{caves,caves2}.  Resolution scales less
than linearly with energy, but probe states~can be constructed with
macroscopic amounts of energy.

\section{Distinct events in spacetime}

We have studied distinct change allowed by shifts of a classical space
or time coordinate.  This symmetric treatment of space and time extends
naturally to spacetime, with an inertial reference frame modeled as a
uniformly shifting classical coordinate system.

\subsection{Distinctness defines relativistic energy}

We consider only bounds in flat spacetime
(\cf\cite{lloyd-nature,brown}), where equivalence of relativistic
quantum field theory to unitary evolution in a Hilbert space with a
finite \mbox{number} of degrees of freedom---the case we have
analyzed---is well established~\cite{weinberg}. This equivalence is
easily demonstrated if spatial resolution is assumed to be finite and
the total size of the system is also assumed finite.  These assumptions
seem unavoidable for quantum field theory to be well defined
mathematically~\cite{duncan} and they underlie the use of finite
lattice models of fields---which make \mbox{accurate} and
systematically improvable predictions~\cite{creutz}.

Now, in flat spacetime, the relativistic ground state (\textit{the
  physical vacuum}) must have energy zero. This is required by frame
invariance~\cite{wightman,wald}.  As we have seen, only energy above
the ground state can cause distinct change within a quantum system, so
average relativistic energy $E$ in any frame is the total energy that
can cause distinct change.  For an evolution that traverses many
distinct states, with $E_0=0$, \eqn{etau} becomes $1/\tau \le 2 E/h$.
In natural units with $h=2$, this is simply
\begin{equation}\label{eq.energy}
1/\tau\le E\;.
\end{equation}
{\em Total relativistic energy $E$ is the maximum average rate of
  distinct state change in a frame, so $E\ps\T$ is the maximum number
  of distinct events possible in time $\T$}.

\subsection{Distinctness defines energy of motion}

We are free to divide macroscopic relativistic energy $E$ into
different forms of energy, and correspondingly partition the total
maximum rate of distinct state change into a sum of different kinds of
change.  For example, for a system with a rest frame energy $E_r$, the
energy $E-E_r$ is the maximum rate of distinct change due to not being
at rest.  This is the conventional {\em kinetic energy}.

Kinetic energy does not, however, provide a natural relativistic
division of distinct change into that seen in the rest frame, plus
additional change seen only in a moving frame
(Figure~\ref{fig.moving-p}).  The quantities $E$ and $E_r$ are maximum
rates in two different frames, and so their difference is not a rate in
either frame.  The natural relativistic measure of extra distinctness
possible in a moving frame is
\begin{equation}\label{eq.vp}
E-E_r/\gamma=v p\;,
\end{equation}
which instead subtracts a rate of distinct change in the rest frame
\textit{as seen in the lab frame.}  From our earlier discussion of
shifts we recognize $v p$ as the average energy of a constant rate
shift $\op{H}_{\text{shift}}=\vec v \cdot {\vec{\op{p}}}$, with $\vec
v$ in the direction of average momentum.  Frame motion is such a shift.

{\em Shift energy $v p$ is a portion of the total energy $E$ that
  allows additional distinct state change in a moving frame, shifting
  at speed $v$, that is not visible in the rest frame.}

\subsection{Distinctness defines momentum}

The appearance of shift energy $v\ps p$ in \eqn{vp} reflects the
fundamental division of special relativistic evolution with $v<c$ into
two parts: rest frame dynamics plus shifting frame motion.  All
state-change except that due to overall motion can be seen in a frame
with zero quantum average momentum---the rest frame.  This includes all
wavefunction spreading, hence most localization.  Conversely, an
evolution describing \textit{just overall motion} has no rest-frame
state-change at all (like a system with $v=c$).  For its wavefunction
to be unchanging when viewed in the rest frame, all momenta must move
in a single direction with a single phase velocity.  Overall momentum
is then due entirely to this one-dimensional shifting evolution, so the
bounds \eqn{fnp} with a minimum $p_0=0$ apply.

For a system well localized in space, an overall-motion wavefunction is
well defined and has a large number of effectively distinct shifts.
Again using units with $h=2$, the momentum analog of \eqn{etau} gives a
bound like \eqn{energy}:
\begin{equation}\label{eq.momentum}
1/\lambda_{\text{motion}}\le p\;.
\end{equation}
{\em Momentum $p$ is the maximum average spatial rate of distinct
  change due to motion, $p\ps\X$ the maximum number of states distinct
  due to a motion of length $\X$, and $p \ps v$~the maximum per unit
  time---the energy of distinct motion.}

\subsection{Invariance counts rest-frame events}

Consider an isolated well-localized system that undergoes a long
evolution that is maximally distinct for both energy and momentum.
Then bounds \eqn{energy} and \eqn{momentum} for shifts in time and
space are achieved, making the identity
\begin{equation}\label{eq.invariance}
E\ps\T - p\ps\X =E_{\text{rest}}\ps\T_{\text{rest}}\;
\end{equation}
a relationship between counts of distinct events with and without frame
motion.  With motion, $\T$ is the time between the starting and ending
events, $\X$ is the distance between them, and $E\ps\T$ is the total
number of distinct events, of which $p\ps\X$ are distinct due to frame
motion.  The difference, $E_{\text{rest}}\ps\Delta t_{\text{rest}}$,
counts the events \textit{not}~due to frame motion.  We can also look
at this in terms of energy: if total energy, and hence all forms of
energy, achieve the maximum rate of distinct change, then multiplying
\eqn{vp} by $\T$ gives the relationship \eqn{invariance} between
counts.

This suggests that the quantum evolution that best approximates
classical mechanics is as distinct as its average energy allows; we
confirm this in Section~\ref{sec.classical}.  Then \eqn{invariance}
becomes a relation between the actual counts of distinct events
underlying classical mechanics. Relativistic action $L\,dt=(H-\sum_i
p_i\,v_i)\,dt$ for a system of interacting particles similarly counts
events \textit{not} due to particle motion.  Dynamics over short time
intervals follows a path where this count is
\textit{greatest}~\cite{gray}.  This is the path of most distinct
events in rest frames (\textit{maximum aging} \cite{taylor}) and, since
energy is conserved, \textit{least} distinct particle-motion.



\setcounter{secnumdepth}{1}

\section{The Most-Classical Limit}\label{sec.classical}

A real physical system cannot be more distinct than its energy allows,
so the \textit{infinitely distinct classical limit} only approximates a
\textit{maximally distinct achievable limit}.  In this limit every unit
of energy causes its share of distinct change, which we classically
count as \mbox{located} where the energy is.  Continuous time evolution
becomes effectively discrete and finite state, enabling new kinds of
models and analysis, simplifying the interpretation of classical models
as approximately quantum, and advancing the study of informational
foundations for mechanics.


\begin{figure}[t]
    \includegraphics[height=.335\columnwidth]{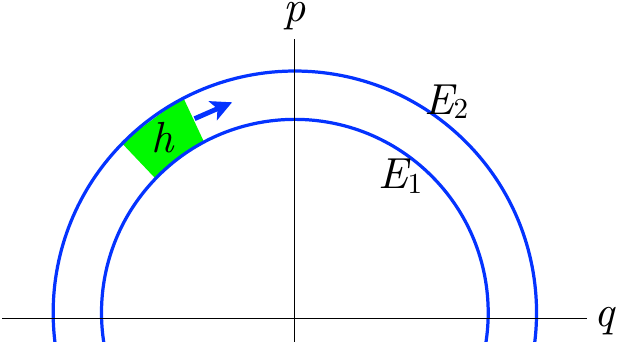}
    \caption{{\it Distinct change in classical phase space}.  For a
      simple harmonic oscillator with energy $E=p^2 + q^2$, all points
      in phase space move on circles of constant energy, rotating once
      every period $T=\pi$.  A minimum distinct area $h$ moves to an
      entirely distinct region in the minimum time $t=h/(E_2-E_1)$: in
      time $t$, $r=\smash{\sqrt{E}}$ sweeps area $\pi r^2\, t/T = E t$,
      so $E_2t-E_1t=h$.}\label{fig.sho} \end{figure}

\paragraph{\bf Classical systems are maximally distinct.}
We can regard classical mechanics as the infinitely distinct $h\to0$
limit of a unitary quantum evolution, in which all amplitude lies on a
single path with stationary classical action
\mbox{\cite{dirac,feynman}}.  A real physical evolution, though, is
only finitely distinct.  Here we show the classical limit approximates
a \textit{maximally distinct evolution}
\mbox{(\cf\cite{adolfo,okuyama,wu})}.  Our analysis treats change as
classically local.  Appendix~\ref{appendix.large} shows that the
underlying quantum change is actually non-local, but can be
consistently modeled as local.

Now, the number of distinct states possible in a given volume of
classical phase space is a fundamental quantity in statistical
mechanics~\cite{peliti,sackur-tetrode}.  For a \textit{single-particle}
state, the momentum bandwidth analog of \eqn{de-t} for large~$N$,
$(p_{\max}-p_{\min})\,\lambda \ge h\ps$, determines the number of
distinct positions for a maximally distinct wavepacket.  For shifts
along each spatial dimension, one distinct position per area $h$ of
{\em momentum-range}~$\times$~{\em spatial-length} is achievable.  This
lets us estimate how many distinct \textit{many-particle} states are
possible with given macroscopic constraints.

Incompressible flow in phase space is the signature that distinct
change due to motion happens as fast as energy bounds allow.  Consider,
for example, the classical phase space of a simple harmonic oscillator
shown in Figure~\ref{fig.sho}.  The evolution of a state (point) is a
circle of constant energy.  A distinct area $h$ {\em crosses a line}
joining energies $E_1$ and $E_2$ in time $h/(E_2-E_1)$: all points move
from one distinct area to another in the minimum time allowed by
\eqn{de-t} for a distinct change in a long evolution with that range of
energies. Any single degree of freedom system similarly achieves this
bound~\cite{haggard}: Hamilton's equations require that an
infinitesimal flow of area $dq\,dp$ crossing an energy gradient $dE$ in
time $dt$ obeys $dq\,dp=dE\,dt$, and integrating this for some time
$\T$ along a line joining $E_1$ and $E_2$ gives $h=(E_2-E_1)\,\T$.
This extends to any number of degrees of freedom: in the simplest case
flow is locally uniform, so all change can be attributed to a single
degree of freedom as a distinct area moves distinctly.

Thus classical mechanics approximates a maximally distinct, hence
effectively discrete, quantum evolution.  Energy bounds are achieved,
so different forms of energy count different kinds of change.  Each
distinct change is like a unit of flow in phase space, where tiny
changes at many locations add up to a distinct amount
(\cf\cite{margolus-pqc}).

\begin{figure}
  \includegraphics[height=.335\columnwidth]{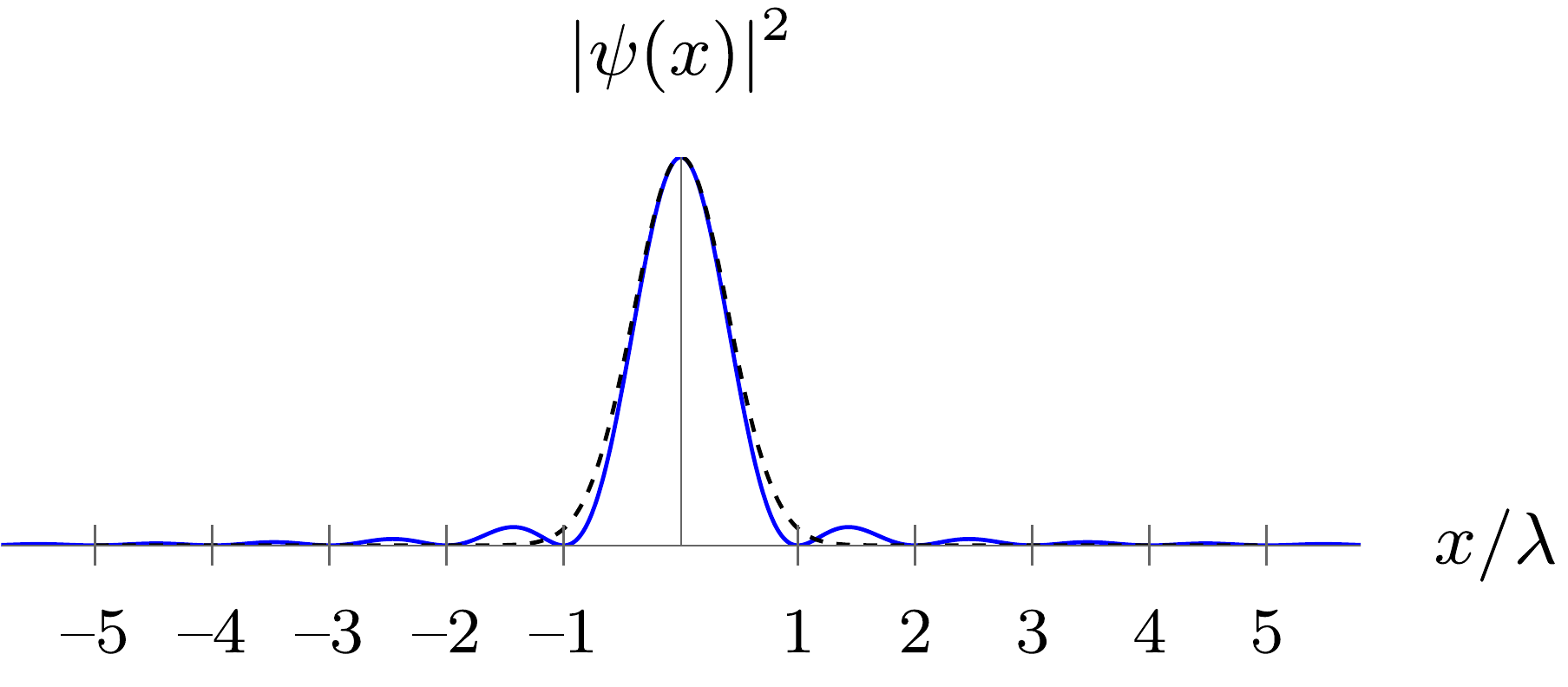}
  \caption{{\it Most-distinct vs Gaussian.}  A wavepacket \mbox{maximally}
    distinct under shifts in space has a probability distribution
    {$\av{\psi(x)}^2$} (solid) similar to a Gaussian distribution
    (dashed).  Nevertheless, standard deviation assigns it an {\em
      infinite} width.}
\label{fig.sinc-vs-gauss}
\end{figure}

\paragraph{\bf The most-classical wavepacket is not Gaussian.}
A Gaussian wavepacket has the minimum product of standard deviations
$\Delta x\ps\Delta p$ allowed by the Heisenberg-Kennard bound
\cite{kennard}.  For this reason it is often taken as the most
classical state describing both location and momentum of a particle.
Modeling classical systems as {maximally} distinct challenges this
idea.  Given any natural measure of momentum width $\Delta p$, distinct
shifts of a wavepacket are at least $\lambda\approx h/\Delta p$ apart.
This is achieved by the widest evenly weighted range of momenta that is
compatible with $\Delta p$---\textit{not} by a Gaussian.

This is illustrated in Figure~\ref{fig.sinc-vs-gauss} for a maximally
distinct $\sinc$ wavepacket (solid) in unbounded space.  Separation
$\lambda$ between distinct shifts determines the bandwidth and hence
the shape.  Such {\em sinc waves} are basic elements of interpolation
theory (see Appendix~\ref{appendix.interpol}).  They are similar to a
Gaussian distribution of the same height (dashed).  Using standard
deviation to measure width in space, as the Heisenberg-Kennard bound
does, a $\sinc$ packet with maximum spatial distinctness and $90\%$ of
its probability between $-\lambda$ and $\lambda$ is assigned an
\textit{infinite} width $\Delta x$.

Gaussian wavepackets provide only a fuzzy bound on distinctness, given
a range of allowed momenta and positions in classical phase space;
uniform bandwidth states achieve maximum distinctness
(\cf\cite{wannier}).  They are also maximally distinct in time for free
motion described relativistically as a uniformly shifting wavepacket.

\paragraph{\bf Some classical models are fundamental.}
It is well known that classical lattice gases, such as the Ising model,
can also be regarded as quantum models~\cite{ruelle}.  These are simple
finite-state models of thermal systems that capture the finite
distinctness that defines entropy.  They also exhibit realistic phase
change behavior, with the same critical exponents as real physical
systems~\cite{ising}.  In view of the effective discreteness and finite
distinctness of classical systems, simple finite-state models of
classical mechanics acquire a similar
status~\cite{hpp,bbm,bbmca,toffoli-diff,fhp,super,wolfram-lga,lgm,rothman,crystalline,ssm}.
Maximum distinctness defines their energies and momenta~\cite{fscm}.

\begin{figure}[t]
  \includegraphics[height=.335\columnwidth]{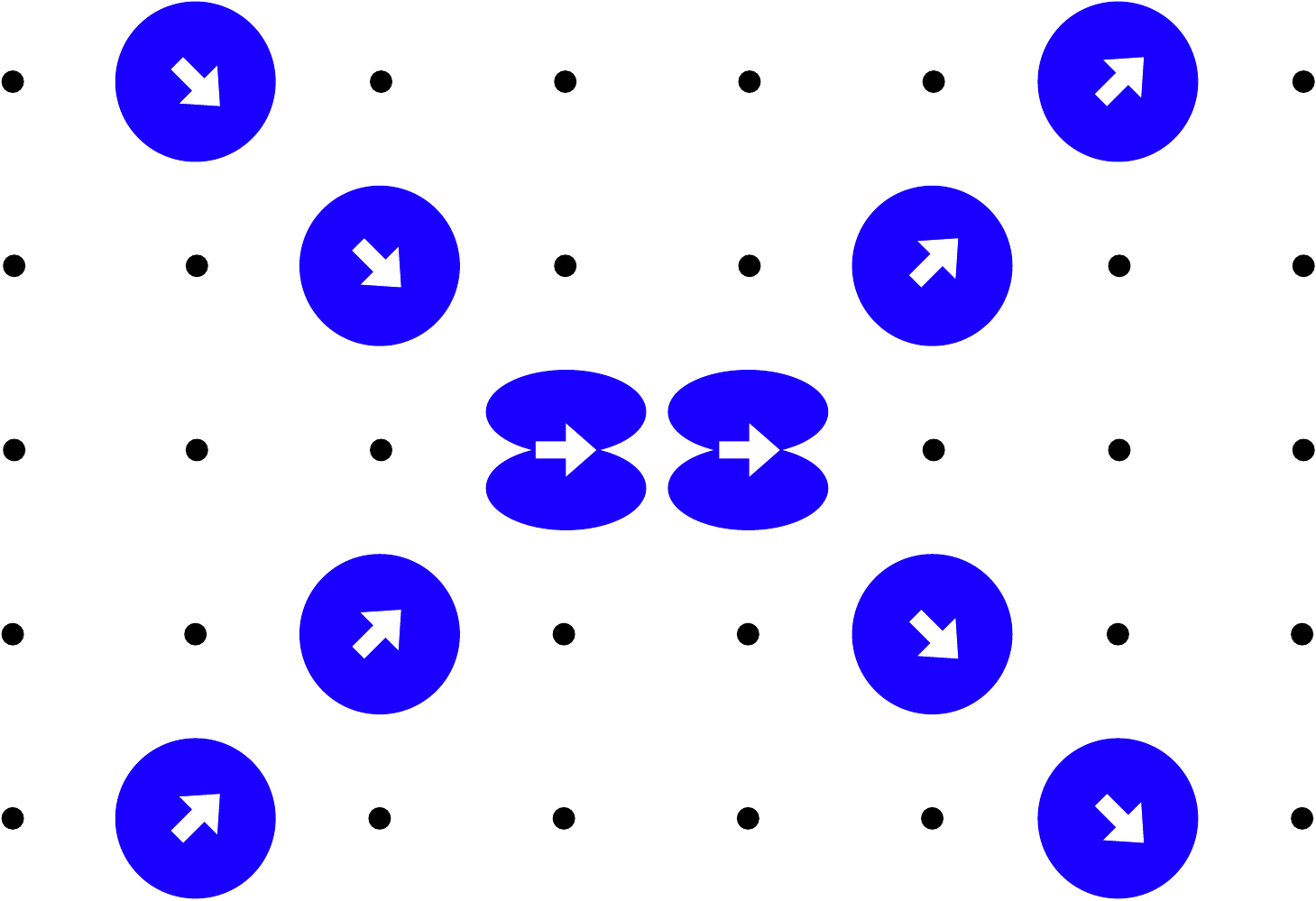}
  \caption{\textit{Lattice gas discretely simulates continuous
      collision}.  Two streams of elastically colliding balls are
    shown. Wavefunction evolution can perform this 2D classical
    computation, continuously interpolating between the discrete ball
    positions.}\label{fig.ssm} \end{figure}

Consider, for example, the finite-state evolution shown in
Figure~\ref{fig.ssm}.  This {\em classical-mechanical lattice gas}
discretely samples a continuous 2D classical evolution~\cite{ssm}.
Here two streams of elastically colliding balls are shown at one moment
of time.  The continuous dynamics is contrived so that, started from a
\textit{perfectly constrained} initial state, the evolution is
equivalent to a finite-state lattice computation at integer times.  We
infer the energy of the lattice gas from the momentum and speed of the
particles when they are moving freely.  Assuming bound
\eqn{momentum}~is always achieved classically, the momentum required to
have distinct particle positions $\lambda$ apart is $p=h/2\lambda$.  If
the distinct positions are $\tau$ apart in time, $v=\lambda/\tau$.
Then relativistic energy is $c^2 \ps p/v$, equal to $h/2\tau$ if
$v=c\,$: all change is motion for massless particles~\cite{mass-note}.
Momentum is defined by $\lambda$, so energy is defined by~$\tau$. This
model is equivalent to a \textit{quantum field theory}, with
most-classical wavepackets representing the particles (see
Appendix~\ref{appendix.field-theory}).

Other lattice gases can similarly be constructed by devising a
continuous classical dynamics plus constraints on the initial state
that give a finite-state evolution at discrete times.  The continuous
dynamics can be somewhat stylized: it may, for example, let classical
particles sometimes pass through each other without interacting.  Such
models enable, for example, discrete molecular-dynamics simulations of
hydrodynamics and complex
fluids~\mbox{\cite{fhp,super,wolfram-lga,lgm,rothman,crystalline}}.  As
in Figure~\ref{fig.ssm}, the models retain exact conservations of the
continuously symmetric dynamics they sample, but lattice constraints
reduce the symmetry of sampled states.  If the lattice has sufficient
discrete symmetry, though, continuous symmetry is recovered in
macroscopic evolutions.  Treating free motion in such models as
maximally distinct defines intrinsic momenta and energies.

\begin{figure}
  \includegraphics[height=.3\columnwidth]{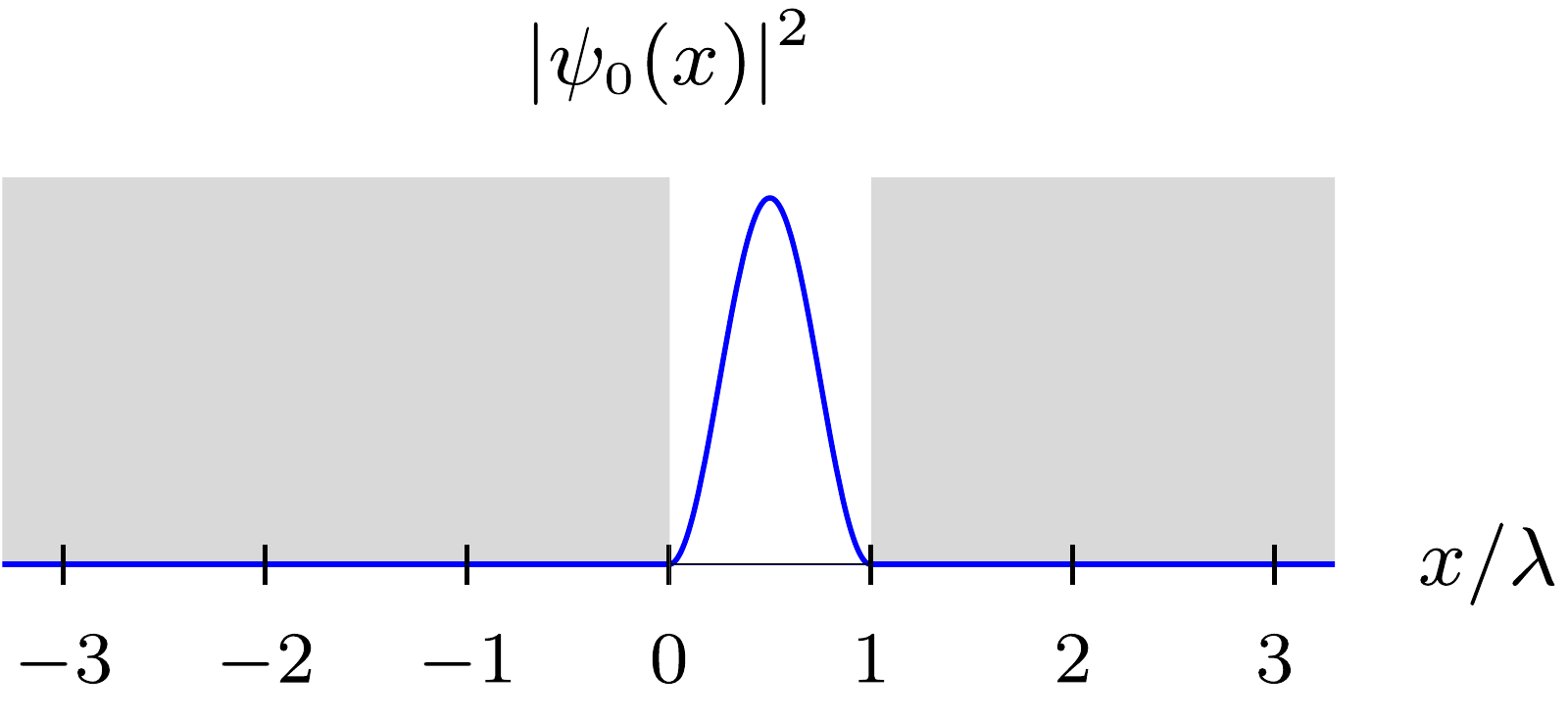}
  \caption{{\it Some energy is tied up in defining system location.}
    For an infinite square well of width $\lambda$, shifts of its
    wavefunction by $\lambda$ are distinct.  Energy $E_0$ used to
    define the well position in space is not available for particle
    dynamics within the well.}
\label{fig.square-well}
\end{figure}

\paragraph{\bf Ground state energy is not uniquely quantum.}  We
define an isolated system by identifying a set of degrees of freedom
that evolve independently of everything else---to some level of
approximation.  If the dynamics is described as wavefunction evolution
with a positive frequency spectrum, its structure imposes constraints
allowing less distinct change than energy (average frequency) would
seem to permit.  For example, in the infinite square well of
Figure~\ref{fig.square-well}, part of the energy of the wavefunction is
used to define the position of the well in space, rather than the
position of the particle within the well.  We can estimate the part
$E_0$ that defines well-position using distinctness bounds.  Clearly
repeated shifts of the wavefunction by the well-width $\lambda$ must be
distinct, so for a wavefunction with mean momentum zero, \eqn{finf} and
\eqn{fnp} require energy
$\ns\langle\ps\op{p}^2\ps\rangle/2m\ns\ns\ge\ns\ns h^2/24 m \lambda^2$
and $\bk{\ps\av{c\,\op{p}}\ps}\ge h c /4\lambda$.  Exact minima are a
bit larger because well-position is so sharply defined ($\psi_0$ is
shown).  Similarly, from \eqn{etau} with $N= 2$, the minimum average
energy $E$ above the physical vacuum for any isolated dynamics to have
a distinct period $T$ is that of a simple harmonic oscillator,~$h/2T$.

Energy required to define the structure of a system is not available to
cause distinct change within it.  This property may
be relevant to the question of whether vacuum energy gravitates in
general relativity \cite{volovik,jacobson-private,kempf-vacuum}.

\paragraph{\bf Even \textit{classical} unitary evolution is
  probabilistic.} Unitary transformations preserve the length of
vectors.  Thus the magnitudes squared of components of a normalized
vector always add up to one, so can play the role of probabilities in
our definition of widths and in counting distinct states of
transformations.  But are these \textit{really} probabilities, even if
they describe a classical evolution?

Ordinary probabilities represent ignorance about information we could,
in principle, know.  In quantum mechanics, probability amplitudes play
a different role: they define which information exists, and which does
not.  We interpret their magnitude squared as a kind of probability by
formally treating non-existent information as if it were merely
unknown~\cite{marg-mech}.  For example, consider use of the occupation
number basis in quantum field theory~\cite{ziman}.  These basis states
eliminate spurious labels for identical particles by evenly superposing
equivalent labeled states.  This represents the non-existence of labels
by assigning equivalent states equal ``probabilities''.  Non-existence
is not ignorance, so this contributes no entropy.  This works just as
well for classical evolutions (see
Appendix~\ref{appendix.field-theory}).

Similarly, in a maximally distinct evolution, a range of frequencies is
evenly weighted (Figure~\ref{fig.numin-numax},
Proposition~\ref{prop.7}).  Treating the \ans{} as ``probabilities'',
there is no information about which allowed frequency the system has,
and maximum about its Fourier conjugate. In a classical evolution this
has nothing to do with measurement, only with what information exists
(\cf\cite{entropic-uncert,birula,max-entropy,emulation}).

\paragraph{\bf 
Energy is maximum rate of \textit{information change}.} Special
relativistic energy counts all possible distinct events per unit time
without requiring any knowledge of the underlying degrees of freedom.
This allows us to place fundamental bounds on classical information.

For example, with $h=2$, relativistic energy is not only the maximum
rate of \textit{distinct change} for a long evolution, it is also the
maximum rate of \textit{classical information change}
(\cf\cite{brem-comm,bek-comm,deffner-comm}): each fastest distinct
change alters a single bit.  To see this, recall that a fastest
evolution is effectively discrete, defined by an evenly spaced series
of distinct states that form a basis for it (see
Appendix~\ref{appendix.interpol}).  Each transition between two
consecutive basis states is a one bit change in time, and we can number
the states with a classical label that reflects this \cite{gray-code}.
To identify energy with the rate of change of the overall state, each
local unit of energy must cause its share of each global change, but
classically we count the changes as located where the energy is (see
Appendix \ref{appendix.large}).  Each distinct local change is thus a
\textit{one bit tick} of a global time count.

This \textit{change bound} lets us estimate how many classical bits
\textit{can be stored} with energy $E$ in a region of radius~$R\ps$. If
fastest change lets all energy leave region in time $2R/c$,
\begin{equation}\label{eq.bits}
  \textit{max change} = E \times 2 R/c \textit{ bits}
\end{equation}
may leave: slightly less than Bekenstein's bound
\cite{bek-ratio,ivanov,casini,longo}.

\begin{figure}
  \includegraphics[height=.3\columnwidth]{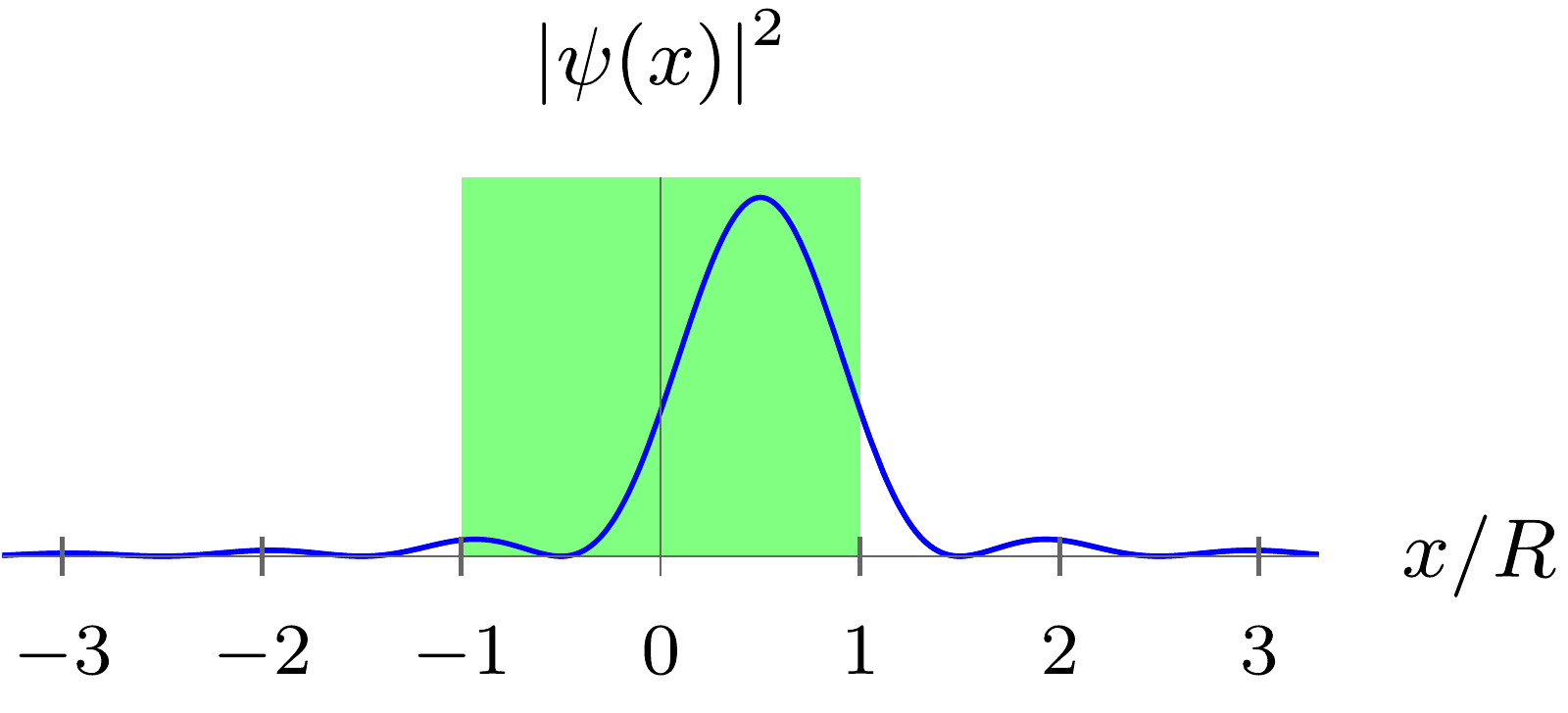}
  \caption{{\it Minimum energy per bit.}  This massless particle has
    the least energy $E$ to have two distinct positions in a region of
    radius $R$ (green) \textit{and} a distinct left or right motion:
    2~bits.  Min momentum $p=1/R$, min $E= c\ps p$, so min
    $E/\textit{bits}= c/2R$.}
\label{fig.sinc-region}
\end{figure}

It is not surprising that bounding the size of a region bounds the
energy needed to have a bit there since, from \eqn{momentum}, having
distinct positions $R$ apart for a moving particle
(Figure~\ref{fig.sinc-region}) requires momentum $p=1/R$, hence at
least energy $E=c\ps p$ (with no rest-mass contribution).  Then $R$ is
the radius of a region with two distinct positions \textit{and} the
motion is either left or right, agreeing with $2 E R/c=2$ bits
\cite{p-and-q}.  If instead we widen the particle to fill the region,
only the direction bit would reside there.  Then $p=1/2R$ and $E=c p$,
so $2ER/c=1$ bit.

\paragraph{\bf Classical spacetime is effectively discrete.} The Planck length and time estimate {\em minimum separations} between pairs of
distinct states~\cite{hoss}. Limits to proximity are expected because
high energy is needed to strongly localize a state in space or time,
but too much energy creates a black hole, making a region inaccessible.
This suggests continuous evolution may be effectively discrete at the
Planck scale~\cite{kempf2,kempf3}, with finer scales inaccessible.

If we model classical mechanics as maximally distinct, however, {\em
  all evolution is effectively discrete}.  Distinct separations $h/2E$
in time can be arbitrarily small for very large objects, but this is
just a property of aggregated change: if enough clocks tick, average
time between ticks can be arbitrarily short.  Given the energy
density~$\rho$ in a region, we can identify the scale $\tau$ of
\textit{individual ticks} there. As above, the energy needed for
distinct motion is least if the motion is at the speed of light, so
distinct positions are $\lambda=c\ps\tau$ apart.  Then, with $c=1$ and
$h=2$, $\rho\,\lambda^3=E=p=1/\lambda$, since distinct volume
$\lambda^3$ must move a width $\lambda$ to move distinctly, and so
\begin{equation}\label{eq.spacetime}
\lambda=\rho^{-1/4}\;.
\end{equation}
Only if $\rho=$ \textit{Planck density} is $\lambda=$ \textit{Planck
  length}.  Effective discreteness links classical \mbox{spacetime} and
quantum dynamics
(\cf\cite{bahr,lloyd-geom,rovelli,mauro,gorard,joe,carroll, geo-qm})
and allows analysis to be performed discretely (see
\cite{kempf1,kempf2,kempf3,kempf4,tsang,marg-mech,emulation} and
Appendix~\ref{appendix.analysis}).

\section{Conclusions}

The standard quantum description of change is partly classical.
Unitary time evolution depends on a quantum description of energy and a
classical length of time.  A shift of a state in space depends on a
quantum description of momentum and a classical length of shift.  Other
single-parameter unitary transformations of a quantum state are
similar: they depend on a quantum observable and a classical parameter.
Thus a fundamental question is, How many distinct (mutually orthogonal)
quantum states can occur in a {\em classical length} of transformation?

We can answer this question in general, given only an {\em average
  width} of the observable in the state it transforms.  For natural
definitions of width, the maximum number of distinct states is
approximately $\textit{length}\times\textit{width}\,/\,h$, and the
exact number is easy to calculate.  This result unifies counting of
distinct states in statistical mechanics and~in dynamics:
momentum-widths count distinct states per unit distance, energy-widths
count them per unit time.  Classical momentum and energy are
essentially widths.

We can view these bounds as a new kind of uncertainty relation that
links classical and quantum distinctness:
\textit{classical-length-per-distinct-state} $\times$
\textit{quantum-width}~$\gtrsim h$.  As in traditional uncertainty, the
minimum product is defined by Fourier complementarity of the length of
a change and the width of its cause.  Here, though, \textit{both} are
always well defined, since a change has a classical length and a
quantum cause.  States that achieve the minimum product resemble
minimum uncertainty states, but are actually elements of interpolation
theory.

Physical distinctness is always finite.  \textit{Infinitely distinct}
classical mechanics approximates a {\em maximally \mbox{distinct}}
quantum world, with \textit{classical energy} identified with the
\textit{actual rate} of distinct state change.  Since classical energy
is essentially local but distinct change is global, equating the two
implies correlation.  For every local unit of \mbox{energy} to cause
its share of each global change, isolated systems must be
\textit{energetically entangled}.  Classically we treat the isolated
systems as unentangled and instead count each system's energy as
\textit{local change} to get the same total rate.

Modeling classical mechanics as maximally distinct dramatically
simplifies its quantum analysis.  If \mbox{classical} energy counts
distinct change, then different forms of classical energy count
different kinds of change.  Special relativity relates counts of
distinct events with and without overall motion, relativistic
Lagrangians count events in rest frames, and dynamics locally maximizes
this count of \textit{proper aging}.  Special relativistic energy is a
conserved total rate that includes all forms of distinct change at all
scales, and so defines \textit{fundamental limits} for information
transformation, transmission and storage: each distinct change alters
\textit{just one bit} of classical information.

Any maximally distinct unitary evolution is \textit{effectively
  discrete}: it is equivalent to a discrete sequence of basis states,
with intermediate states interpolated from them.  This makes
approximately-classical spacetime effectively discrete at a scale
\textit{set locally by energy density}, and makes maximally-distinct
field theories equivalent to discrete-spacetime models.  Conversely,
finite-state lattice models of classical mechanics can be recast as
special-cases of continuous quantum field theories, with rates of
distinct change in spacetime defining classical four momentum.

Finally, finite distinctness exposes the informational foundations of
mechanics.  Modeled as maximally \mbox{distinct} quantum evolution,
classical mechanics becomes an \mbox{effectively} discrete and
finite-state computation in which \mbox{basic} physical quantities,
such as entropy and energy, are also basic computational quantities,
such as memory and {processing} rate.  This provides a simplified
informational context in which dynamics is governed by an interplay of
counts, and concepts such as probability amplitudes and ground state
energy have direct classical significance.

\begin{acknowledgments}
\noindent I thank Gerald Sussman, Tom Toffoli, Hrvoje Hrgov\u{c}i\'{c},
Ted Jacobson, Samuel \mbox{Braunstein}, Lorenzo Maccone, Hal Haggard,
J. Corwin Coburn, Seth Lloyd, Deepak Dhar and Adolfo del Campo for
valuable discussions.  I also thank Sussman and the MIT CSAIL
Laboratory for aiding and encouraging this research.
\end{acknowledgments}

\setcounter{secnumdepth}{1} \appendix

\newcommand{\ft}{{\psi}}
\newcommand{\td}{{\tau_\delta}}

\newcommand{\bn}{{b,\sss N}}
\newcommand{\bi}{{b,\sss\infty}}
\newcommand{\bnt}{{b,{\sss N},\tau}}
\newcommand{\tbn}{{\tau,b,\sss N}}
\newcommand{\tbi}{{\tau,b,\sss\infty}}

\section{Periodic interpolation}
\label{appendix.interpol}

A continuous time evolution that uses only a finite range of Fourier
frequencies is \textit{effectively discrete}: it can be exactly
reconstructed from discrete samples.  For example, if a complex-valued
wave has period $T$, only frequencies $m/T$ for integer $m$ can appear
in its Fourier sum, and only a finite number $N$ can fit into the
finite range.  Then $N$ values of the wave determine all $N$
coefficients in the sum, hence the evolution at all times
(Figure~\ref{fig.finite-bw}).

To construct a set of $N$ distinct (orthogonal) localized waves, let
any one of $N$ \textit{equally separated} distinct wave values be one,
the rest zero (Figure~\ref{fig.packet}).  Each wave is then maximally
localized in time, and so minimally in frequency: the $N$ frequencies
are evenly weighted.  If the distinct times have unit spacing and the
frequency range is centered at $b$, a localized Fourier sum is
\begin{gather}\label{eq.sincn}
\sinc_\bn u\,=\,
  \begin{array}[c]{@{}l@{}}\\[-1.1em]
    {\scriptstyle \phantom{m\,=\,} +\frac{N-1}{2}} \\[-.4em]
    {\displaystyle \sum} \\[-.2em] 
    {\scriptstyle {m\,=\,} -\frac{N-1}{2}} \\[-.1em]
  \end{array}
  \;\frac{1}{N} \,e^{2\pi i u\,(b\,+\,m\ns/\ns N)} \,.\,
\end{gather}
This function of $u$ has period $N$ in both magnitude and phase if $b$
is chosen so the frequency range starts at an integer multiple of
$1/N$.  Always periodic in magnitude, it has {\em magnitude$\,$one} for
integer $u \equiv0\bmod N$, and {\em zero} for other integer $u$. It is
a periodic generalization of the \textit{normalized} $\sinc u$ function
of interpolation theory:
\begin{gather}\label{eq.sincb}
\sinc_{\sss 0,\infty} u\,= \,\int_{ -\frac{1}{2}}^{ +\frac{1}{2}}
d\nu\,e^{2\pi i u \nu}\,=\,\frac{\sin\pi u}{\pi u}\;.\;
\end{gather}



To construct a localized wave with time $\tau$ between its distinct
shifts, let $u=t/\tau$ in \eqn{sincn}, so unit-separated $u$ are
separated by $\tau$ in time.  Then all frequencies $b+m/N$ in the sum
are multiplied by $1/\tau$ and the center frequency becomes $b/\tau$.
Now, consider a periodic function $s(t)$ with a given period $T$ and
frequency range (\ie, a range starting at an integer multiple of $1/T$,
with a known center).  Equality in \eqn{wt} gives the maximum number
$N$ of independent values and average time $\tau=T/N$ between them.  We
can express $s(t)$ as a sum of $N$ sample values $s(n\tau)$ multiplying
waves localized at $N$ sample times $n\tau$:
\begin{equation}\label{eq.sampling}
s(t)=\sum\nolimits_{n=0}^{N-1} \;\sinc_\bn({t}/{\tau}\ps-\ps
n)\;s(n\tau) \;.
\end{equation}
Equality is obvious if $t$ is one of the $N$ sample times, since if
$t=k\tau$ for integer $k$, $\sinc_\bn(k-n)$ is a periodic delta
function: 1 if $k-n\equiv0\bmod N$, 0 otherwise.  The $N$ sample values
determine the $N$ Fourier coefficients for the $N$ frequency
components, hence $s(t)$ at all times.

We can generalize \eqn{sampling} to \textit{any vector evolution}
$\ket{\ft(t)}$ with a given period and frequency range:
\begin{equation}\label{eq.sampling-time}
\ket{\ft(t)}=\sum\nolimits_{n=0}^{N-1} \;\sinc_\bn({t}/{\tau}\ps-\ps
n)\; \ket{\ft(n\tau)}\;.
\end{equation}
Here we interpolate evolution of each component of $\ket{\ft(t)}$ from
its values at equal time separations $\tau$.  This identity does not
require the $\ket{\ft(n\tau)}$ to be distinct vectors, but if they are
evolution is \mbox{\textit{maximally distinct}}
(Figure~\ref{fig.unitary}). If~the $\ket{\ft(n\tau)}$ are
\textit{orthonormal}, $\ket{\ft(t)}$ is {\em normalized at all times}:
$\braket{\ft(t)}{\ft(t)}=\sum_n \sinc^2_{\sss 0,N}(t/\tau\ps-\ps n)=1$.
This sum is just \eqn{sampling} with \raisebox{0pt}[.8em][0pt]{$s(t)=$}
$\sinc_{\sss 0,N}(\vartheta/\tau-t/\tau)$ in the limit
\mbox{$\vartheta\to t$}.  Any $\tau$-separated set of samples of this
$\ket{\ft(t)}$ is similarly orthonormal and forms a basis for this
\textit{unitary} evolution.

\eqn{sampling-time} applies to any periodic quantum evolution
\eqn{kett}.  From Proposition~\ref{prop.7}, a maximally distinct
evolution has a finite equally-weighted range of frequencies, so its
\mbox{center} frequency $b/\tau$ is its average energy $E$ over $h$
(Figure~\ref{fig.numin-numax}).  If the range starts at $E_0/h$, the
evolution is maximally distinct for its average energy and for all
frequency widths compatible with that $E$.  Maximally distinct
evolution is effectively discrete, erasing many distinctions between
continuous and discrete {(see
  \cite{kempf1,kempf2,kempf3,kempf4,tsang,marg-mech,emulation} and
  Appendix~\ref{appendix.analysis})}.



Of course in non-relativistic evolutions we generally \mbox{assume} the
lowest energy (frequency) can have any value, and so centering of $b$
would not be restricted.  Then an evolution periodic in the sequence of
physical states visited would generally give them a different phase
with each repetition.  Even in this case, though, \eqn{sampling}
remains valid: for $t=k\tau$, $\sinc_\bn(k-n)$ acts as a delta function
for the first period, and hence determines the values that get repeated
with a changing phase.  Thus \eqn{sampling-time} also continues to
hold, and differing phases do not affect distinctness.

For example, for maximally distinct evolution between $N=2$ distinct
states with $b=0$ (\ie, $E=0$), \eqn{sampling-time} gives
\begin{equation}\label{eq.fastest}
\ket{\ft(t)}=\cos{\frac{\pi t}{2\tau}}\,\ket{\ft(0)} \,+\,
\sin{\frac{\pi t}{2\tau}}\,\ket{\ft(\tau)}\;,
\end{equation}
which gains a factor of $-1$ for each cycle of alternation. Similar
analysis applies to other single-parameter unitary transformations
(Figure~\ref{fig.unitary-l}).  For rotations of fermions, half-integer
frequencies have a fundamental significance.

\section{Numerical methods}\label{appendix.numerical}

\vspace{-.05em} In order to independently determine bounds and to
verify our analysis, we study minimum frequency-widths computationally
for evolutions with given sets of intervals between distinct states.
This study is based on linear optimization \cite{linear}, which
efficiently finds the global maximum or minimum of a linear function of
a large number of variables that obey a set of linear constraints.  In
this case the variables are the probabilities \ans{}, which obey linear
orthogonality constraints~\eqn{orthogt}.  One linear function we can
maximize, subject to the constraints, is the sum of probabilities in a
range of frequencies.  We can similarly minimize any average deviation
$\Delta\nu=\ps\ps\mom{\nu-\alpha}{\M}$ since, by definition \eqn{dnu},
$(\Delta\nu)^\M$ is linear in probabilities.  To limit the number of
variables in the problem we restrict times to be integers, chosen so
that whatever resolution we require in dimensionless ratios of times is
available.  Then, because of \eqn{nun}, maximum period $T$ is also the
maximum number of distinct phases in \eqn{orthogt}, hence the maximum
number of \ans{} to be determined (more cannot affect orthogonality
nor, from Property~\ref{p4}, decrease width).  Since overall phase also
does not affect orthogonality or widths, we are free to take the lowest
frequency $\nu_0=0$ and only study evolutions periodic in both
magnitude and phase.

Several numerical tests are illustrated in Figures.  The code for these
and an extensive set of other tests is available~\cite{code}.  A first
set of stochastic tests verify that equally spacing the distinct states
of a periodic evolution allows the smallest value for any well-behaved
width.  Figure~\ref{fig.eq-is-min} illustrates this for the width
$\Delta \nu= 2\ps\ps\mom{\nu-\nu_0}{1}$, testing sets of unequal
integer intervals with lengths up to 1000.  The dashed line is
approached as all ratios of intervals approach one; all deviation
widths are shown to have similar limiting behavior.
Figure~\ref{fig.stair} shows that bound \eqn{ceil-q} is also achieved
with equal intervals.  Each dot shows a maximum total probability that
can fit into a randomly chosen width of frequency-range, for an
evolution with randomly chosen intervals between up to ten distinct
states.  Larger widths can hold more probability.

For a portion of evolution with average separation $\tau$ between $N$
distinct states, we find numerically that any width $\W$ is minimized
when the full evolution has period $N\tau$, \textit{as long as this
  periodic bound never decreases with} $N$.  We test this by
considering a full evolution with maximum period ${T\gg N\tau}$.  With
$T$ sufficiently large, the minimization becomes independent of $T$.
We use this property first to verify that equal spacing within the
portion is optimal, then to check bounds as in
Figure~\ref{fig.theory-vs-data}.  Convergence to a global minimum for
large $T$ is directly illustrated in Figure~\ref{fig.width-vs-period};
variation of minima falls off like $T^{-2}$
asymptotically~\cite{unpublished}.  We separately confirm periodic
bounds are {\em exact} by letting $T$ be a large integer multiple of
the minimum bandwidth period $N\tau$, and verifying a few thousand
cases to a thousand decimal digits each.  The maximizations of
Figure~\ref{fig.qmax} similarly use $T\gg N\tau$.  For various values
of bandwidth $\W$, max $q$ is determined.  Unequal spacing within the
portion requires $\W_{\sss q=1}\ps\tau >\ns 1$, which numerically
verifies Proposition~\ref{prop.8}.

\section{Bounds on part of an evolution}\label{appendix.portion}

If a portion of an evolution traverses $N$ distinct states with average
separation $\tau$, the minimum for an average width $\W$ is generally
achieved if the evolution has the least possible bandwidth, hence
period $N\ps\tau$.  We analyze this for deviation widths and explain
their exceptions.

\vspace{.2em}

\prop8 For evolution with bandwidth $\W$, in which a portion has $N$
distinct states an average of $\tau$ apart, $\W\,\tau$ is minimized in
evolutions with period $N\ps\tau$.


\begin{figure}[t]
  \includegraphics[height=.32\columnwidth]{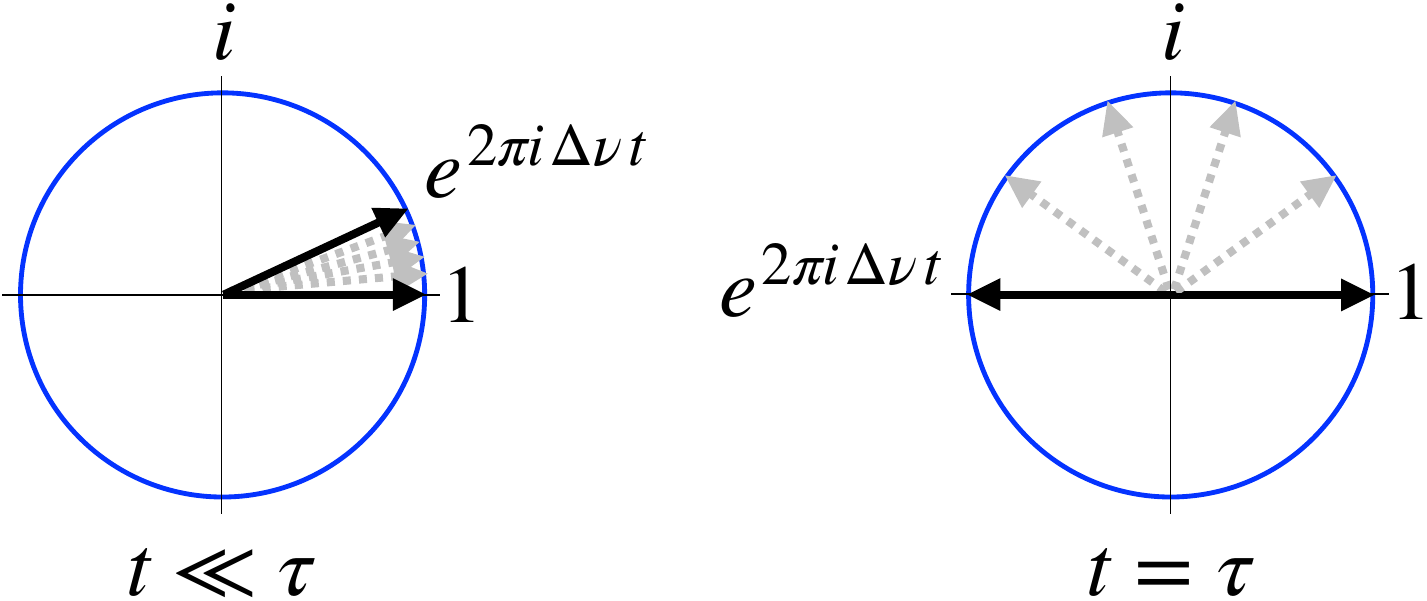}
  \caption{{\it Two frequencies give fastest orthogonality.} The
    earliest that $\braket{\psi(t)}{\psi(0)} = 0$ is when fastest and
    slowest changing phases (black) cancel and there are no other
    phases (gray). Equally weighting max and min frequencies then gives
    zero.}
\label{fig.cancel}
\end{figure}

\proof Consider first the case $N=2$, shown in
\mbox{Figure}~\ref{fig.cancel} with lowest frequency $\nu_{\min}=0$ and
highest $\nu_{\max}=\W$.  We seek the least time between orthogonal
states: when $\braket{\psi(t)}{\psi(0)}= \sum_{\sss n\min}^{\sss n\max}
\av{a_n}^2 \, e^{2 \pi i \nu_n t}\ns=0$.  The earliest time $\tau$ the
sum can be zero is when min and max phases (black) are opposite, with
no other phases (gray) so the imaginary part is zero.  An equally
weighted sum is then zero, period is $2\,\tau=1/\W$, and this smallest
$\tau$ minimizes $\W\,\tau$.  The case $N>2$ is similar, but requires
more frequencies (and phases) with non-zero weight, and so more time
$\tau$ between distinct states than in
\mbox{Figure}~\ref{fig.cancel}. The smallest number of constraints
\eqn{orthogt} and hence frequencies is $N$, achievable only in an
evolution with period $N\ps\tau$.

That least constraint (equal spacing within a portion) allows the
smallest $\W\ps\tau$ is easily verified: with unequal spacing,
$\W\ps\tau>1$ \cite{code,bw1-note}.  With equal spacing, periodic
evolution can achieve \eqn{de-t}:
$\W\ps\tau=(N-1)/N<1$.~\hfill$\square$

Thus any bound on periodic evolution applies also to portions as long
as the width is minimized by a min-bandwidth completion of the portion.
Below we analyze only {\em generalized deviation widths} $\W$.
Exceptions occur when the min-bandwidth distribution for period
$N\ps\tau$ is not centered on one of the $\nu_n$: deviations from
$\bar\nu$ with $N$ even.  Then, for small values of $M$, a slightly
wider set of frequencies allows probability mass to be moved closer to
$\bar\nu$, decreasing $\W$ as long as the penalty for using more
bandwidth is small.  We analyze when this helps.

\prop9 For $N=2$ and $T\ne N\ps\tau$ the narrowest possible
distribution uses three frequencies, and is a unique function of $T$,
symmetric about the mean.

\proof With period $T\ne N\ps\tau$ there are three independent
constraints from \eqn{orthogt} for $N=2$: one for zero separation and
two for separation $\tau$ (real and imaginary parts).  Thus the weights
\ans{} for {\em three consecutive} $\nu_n$ are determined---this is the
narrowest distribution.  For $2\ps\tau < T < 4\ps\tau$ a solution
exists with three \textit{positive} \ans{}$\,$: \mbox{\{$p$, $1-2p$,
  $p$\}}, where $p=1/(4 \sin^2 \pi\ps \tau\ns\ns/\ps
T\ps)$.~\hfill$\square$

If maximum period $T > 2\ps\tau$, we can still have minimum bandwidth
with three consecutive $\nu_n$: let $T=4\ps\tau$, so the frequencies
are twice as close as for $2\tau$. Then $p=1/2$ and the three \ans{}
are \mbox{\{$1/2$, $0$, $1/2$\}}: two equally weighted frequencies at
the spacing for period $2\ps\tau$.

\prop{10} For $N=2$, minimum bandwidth minimizes any
deviation-from-$\bar{\nu}$ width $\W$ for $\M\ge \pi/2$.

\proof The narrowest distribution with $T\ne N\ps\tau$ has a mean that
is the middle of three consecutive $\nu_n$, allowing probability mass
closer to the mean than with two $\nu_n$. To find $T$ that minimizes
$\W\,\tau$ we insert into deviation \eqn{dnu} from the mean the
distribution $\{p,1-2p,p\}$ given above; the derivative of $\W\,\tau$
with respect to $T$ is zero if
\begin{equation}\label{eq.tan}
\frac{2\,\pi \tau\ns/T}{\tan \ps\pi \tau\ns/T} = \M\;.
\end{equation}
For $0<\M<\pi/2$ this has solutions with $2\ps\tau<T< 4\ps\tau$. In all
cases, minimum $\W\,\tau$ is smaller than for $T=2\ps\tau$. For $\M =
\pi/2$, maximum period $T = 4\ps\tau$, actual period is $2\ps\tau$, and
we revert to the minimum bandwidth bound.~\hfill$\square$

\begin{figure}[t]
  \includegraphics[height=.32\columnwidth]{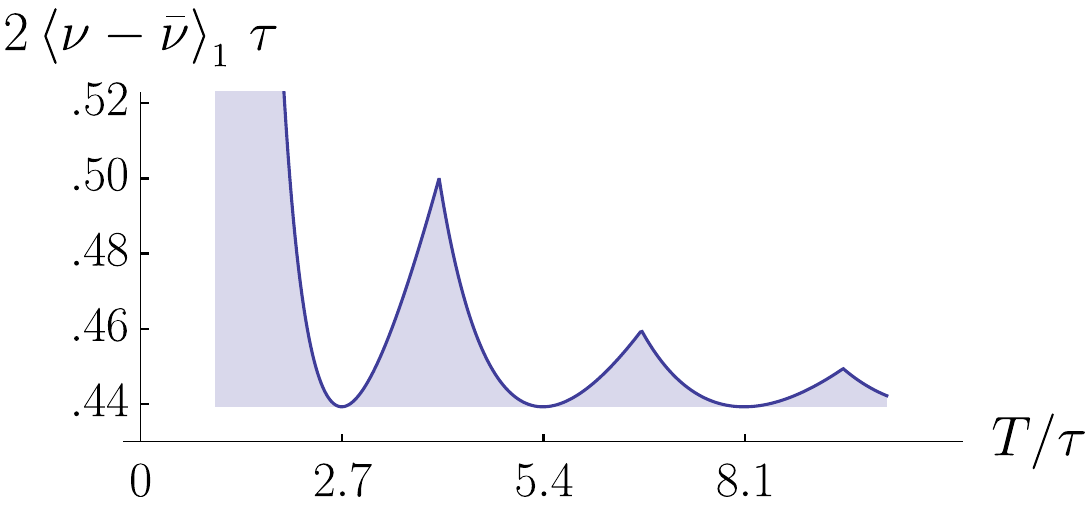}
  \caption{{\it Minimum of $\Delta \nu\,\tau$ as $T/\tau$ varies}.  We
    numerically find the minimum of $2\ps\ps\mom{\nu-\bar\nu}{1}\,\tau$
    for two distinct states with separation $\tau$ in a total period
    $T$, for a range of values of $T$.  The predicted global minimum
    recurs at the predicted $T/\tau$.}
\label{fig.width-vs-period}
\end{figure}

\begin{figure}[t]
  \includegraphics[height=.32\columnwidth]{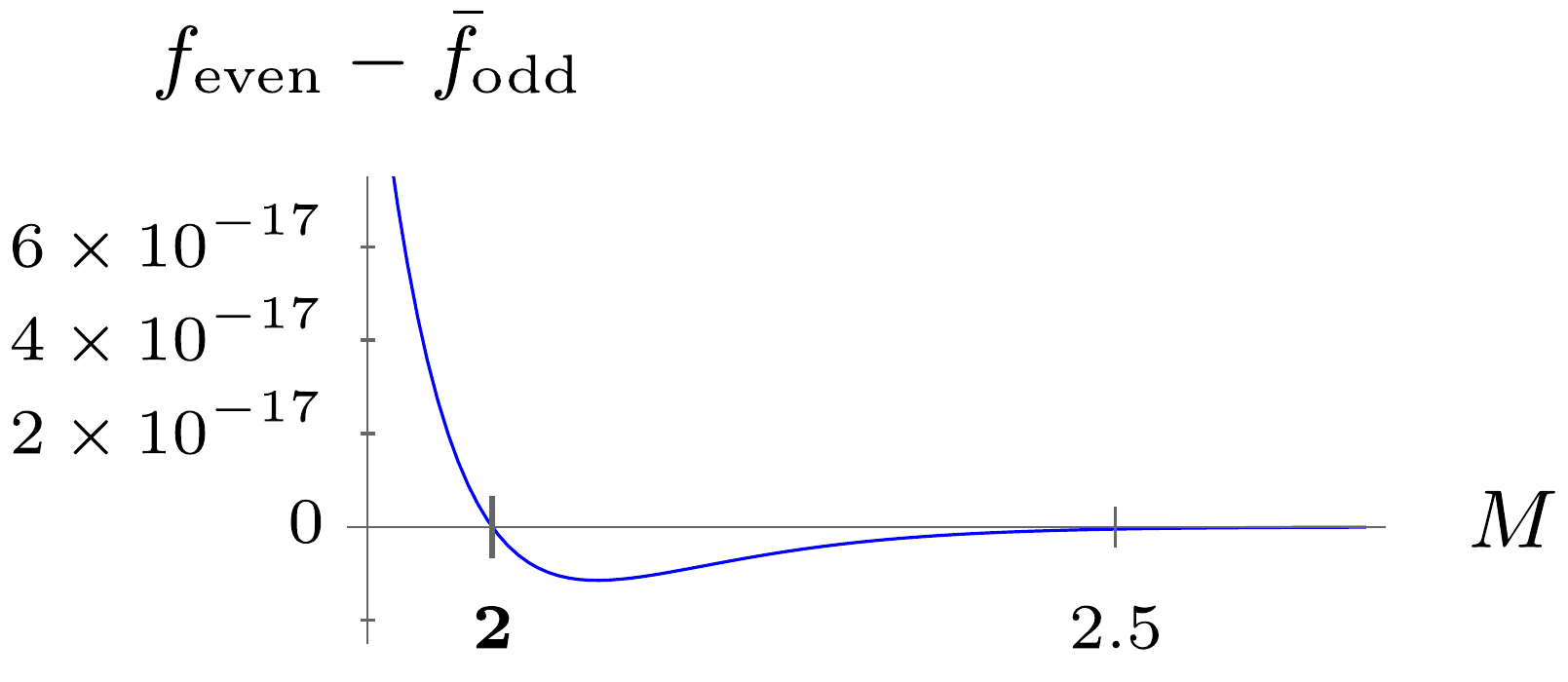}
  \caption{ {\it Threshold for exceptions to minimum bandwidth.}  The
    smallest $\M$ above which the minimum bandwidth bound
    $f_{\bar\nu}(\M,N)$ holds for a portion with even $N$ increases
    with $N$, with a limiting value of 2.  We illustrate this for
    $N=100,000$. For $\M< 2$, $f_{\text{even}} \equiv f(N) >
    \bar f_{\text{odd}} \equiv (f(N+1)+f(N-1))/2$.  The average $\bar
    f_{\text{odd}}$ of always-correct odd-$N$ bounds approximates a
    correct even-$N$ bound, becoming exact as $N\to\infty$.}
\label{fig.threshold}
\end{figure}

For $\M=1$, for example, \eqn{tan} gives $T/\tau=2.69535\ldots$ and
hence $2\ps\ps\mom{\nu-\bar\nu}{1}\,\tau \;\ge\;0.439284\ldots\,$,
which is exactly what we find minimizing numerically (see
Figure~\ref{fig.width-vs-period}).

\prop{11} For $N=2$, minimum bandwidth minimizes any
deviation-from-$\nu_0$ width $\W$ for $\M>0$.

\proof If we insert the distribution $\{p,1-2p,p\}$ into deviation
\eqn{dnu} from $\nu_0$, $\W\,\tau$ always attains its minimum for
$T=4\tau$, the maximum period that keeps $1-2p\ge0$.  This gives a
distribution $\{1/2,0,1/2\}$ and period $2\ps\tau$.~\hfill$\square$

For $N=2$, the bounds above on deviations from $\nu_0$ are known {\em
  speed limits} for all $\M$~\cite{max-speed,luo1,zych}, and known for
deviations from $\bar\nu$ for $\M=1$ and $2$~\cite{mandelstam,luo2}
with extension to $\M>2$ obvious~\cite{mand-note,casella}.  For $N>2$,
analysis similar to the above shows that using slightly more than
minimum bandwidth does not allow smaller bounds as long as the
min-bandwidth distribution is centered on one of the $\nu_n$.  We also
verify numerically that no amount of extra bandwidth helps
(Figure~\ref{fig.theory-vs-data}).  As with $N=2$, the cases not so
centered are widths about the mean with $N$ even, and again slightly
wider bandwidth only helps if $\M$ is small.  We find numerically that
\eqn{fnubar} is an exact bound if $\M\ge2$, which is the threshold
above which \eqn{fnubar} increases with $N$, as it must to be a portion
bound (Figure~\ref{fig.threshold}).


\section{Distinctness of large evolutions}
\label{appendix.large}

\begin{figure}[t]
  \includegraphics[height=.32\columnwidth]{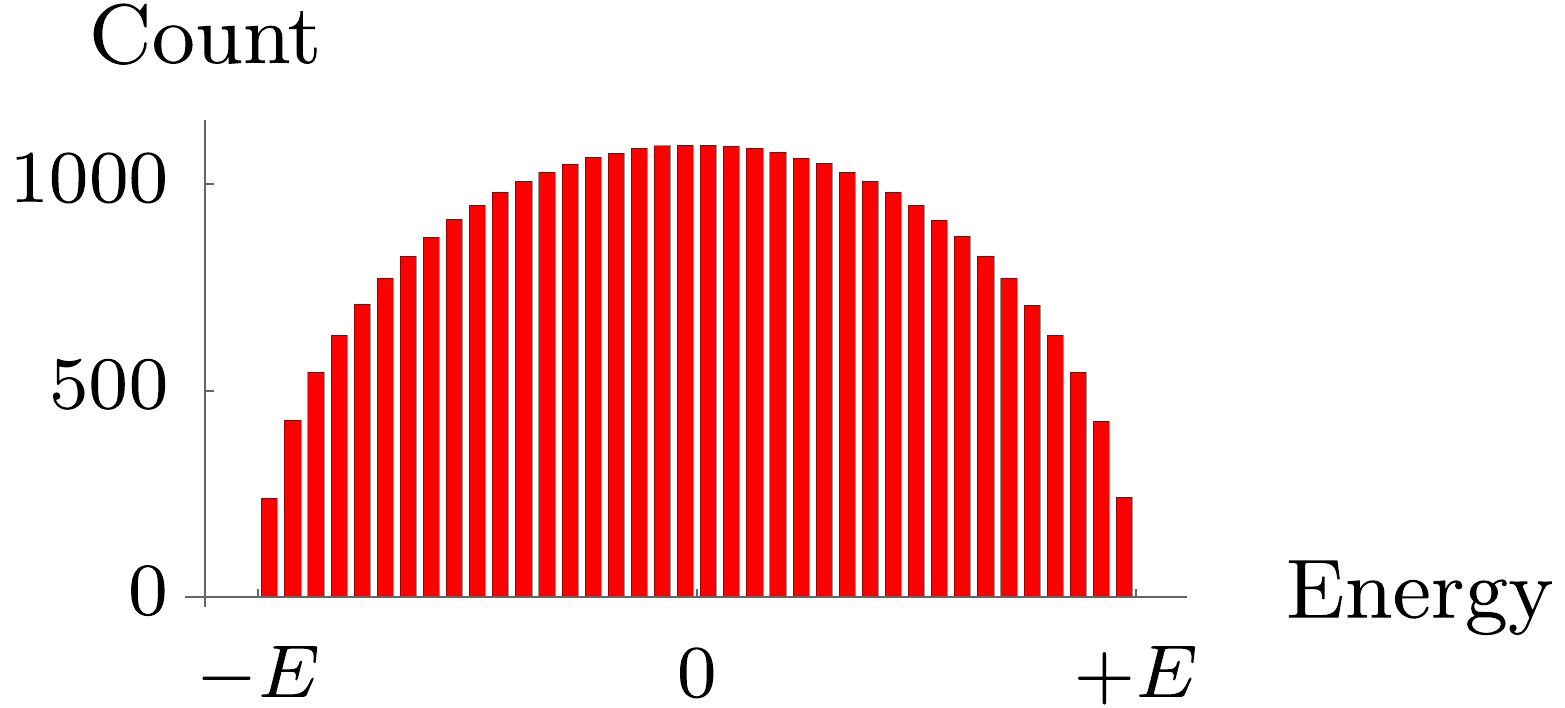}
  \caption{ {\it Eigenvalue distribution for a random Hamiltonian.}  We
    count eigenvalues for energy ranges of an $N\times N$ Hermitian
    matrix with random entries and $N=2^{15}$. The distribution is
    semi-circular for large $N$ regardless of the randomness details.}
\label{fig.random-hamiltonian}
\end{figure}

\newcommand{\EE}{{\mathcal{E}}}

The relationship between complicated Hamiltonian evolutions and
properties of their energy has been studied for evolutions generated by
large randomly-constructed Hermitian matrices~\cite{wigner55}. A
universal property of these is shown in
Figure~\ref{fig.random-hamiltonian} for an $N\times N$ matrix with
$N=2^{15}$.  This is a histogram of counts of eigenvalues in different
ranges for a Hermitian matrix constructed by \textit{adding a matrix},
with independent random complex-valued entries with mean zero and
bounded variance, \textit{to its conjugate transpose}.  The normalized
distribution is semi-circular: $\text{Count}^2\ns/\text{Count}_{\max}^2
+\, \text{Energy}^2\ns/\text{Energy}_{\max}^2=1$ for $N\to\infty$, and
this property is independent of the details of the
ran\-dom\-ness~\cite{code,erdos}.  Thus if $d\EE$ is the width of a
column and $\rho(\EE)\, d\EE$ the fraction of the $N$ eigenvalues in
the energy-$\EE$ column, the continuum \textit{density of states}
\begin{equation}\label{eq.dos}
  \rho(\EE)= \rho(0)\sqrt{1-(\EE/E)^2}\;,
\end{equation}
where ${E\propto\sqrt{N}}$ is the radius of the distribution.  For the
fractions $\rho(\EE)\,d\EE$ to add up to one, $\rho(0)=2/\pi E$.

A {\em generic state} $\ket{\psi(0)}$ of an evolution generated by a
random Hamiltonian inherits the distribution $\rho(\EE)\,d\EE$ for the
\textit{probabilities} of observing different energies~\cite{rmt}.
Then the \textit{average energy} is $E$ above the lowest and
\begin{equation}\label{eq.j1}
  \braket{\psi(t)}{\psi(0)}=\int_{-E}^E e^{\frac{2\pi i}{h} \EE t}
  \rho(\EE)\ps\ps d\EE = \frac{J_1(2\pi Et/h)}{\pi Et/h}\;,
\end{equation}
where $J_1$ is a Bessel function.  The time between zeros is
approximately the minimum $\tau_{\min}=h/2E$, becoming exact for
large~$t$.  This gives a long evolution that is almost
maximally distinct for its average energy.

\begin{figure}[t]
  \includegraphics[height=.32\columnwidth]{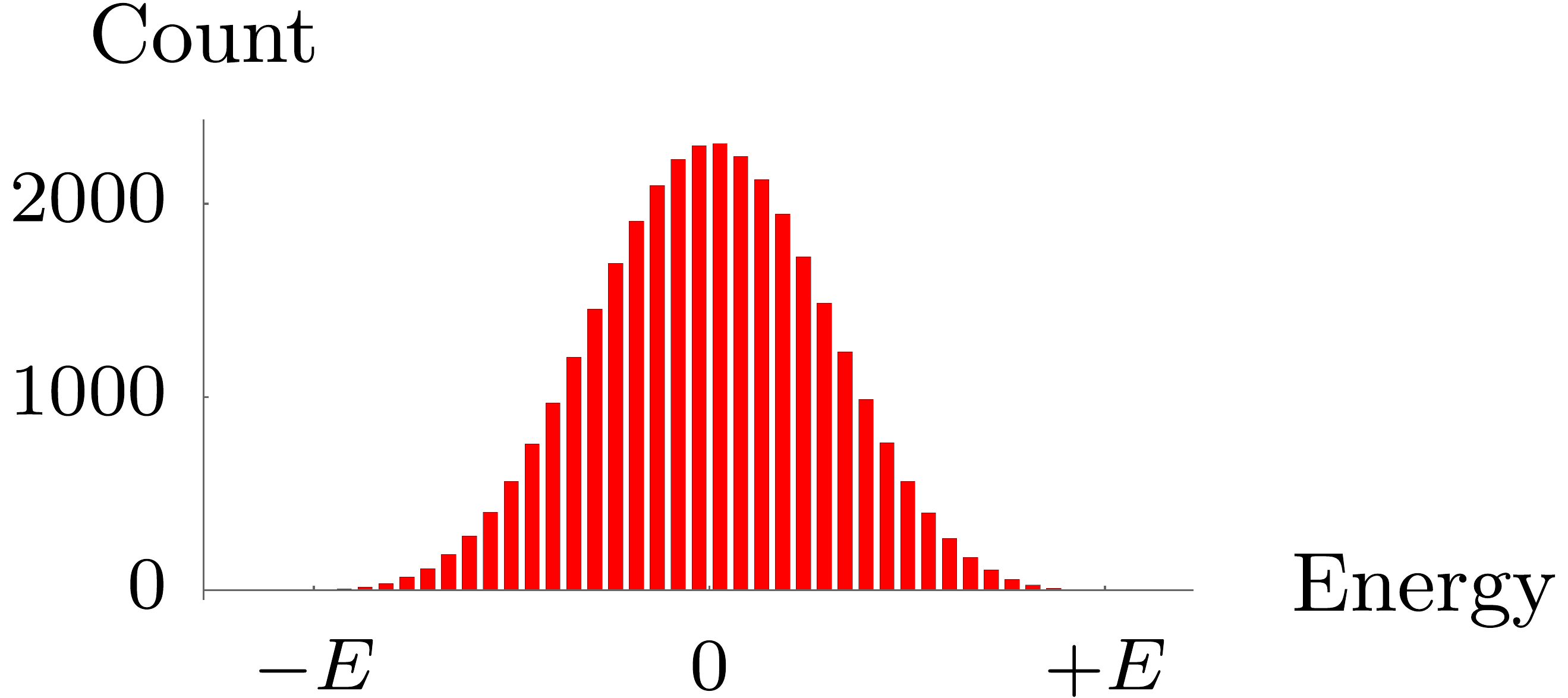}
  \caption{ {\it Distribution for sum of local random~Hamiltonians} is
    a truncated Gaussian.  We count eigenvalues for dynamics of a chain
    of $n=15$ qubits.  The total Hamiltonian is $2^n\times 2^n$, a sum
    of $n$ random Hamiltonians acting on adjacent qubits.}
\label{fig.random-local}
\end{figure}

This analysis does not constrain the dynamics to be \textit{spatially
  local}.  If we do we instead get a truncated normal
distribution~\cite{code,rmt,sum-dist,truncated-normal}.
Figure~\ref{fig.random-local} shows eigenvalue counts in different
energy ranges for a $2^n\times 2^n$ random Hamiltonian acting on a
closed chain of $n=15$ qubits, constructed by summing $n$
\textit{local} random Hamiltonians that have a bounded range of
eigenvalues with mean zero, each acting on only one adjacent pair of
qubits. For large $n$ the distribution's radius ${E\propto n}$ and its
standard deviation $\sigma \propto \sqrt{n}$, so $\sigma\approx
E/\sqrt{n}$.  If we let $\rho(\EE)$ in \eqn{j1} be a truncated Gaussian
with such a $\sigma$ we find, {from a generic start}, the first nearly
distinct state ($\textit{overlap}<\textit{some }\epsilon$) appears at
\raisebox{0pt}[0pt][0pt]{$t\propto\sqrt{n}\, \tau_{\min}\propto
  \sqrt{E}/ \ns E$}, giving a non-extensive rate of distinct change of
order only $\sqrt{E}$ (\cf\cite{kok2}).

This is still not the whole story, though, since we have neglected
\textit{entanglement}.  Consider a large system made up of many
isolated subsystems~\cite{code}.  If these are fully independent, their
tensor product state has a sharply peaked distribution of total-energy
eigenvalues \textit{regardless} of subsystem states, again giving a
total rate of distinct change like $\sqrt{E}\ps$: \textit{unentangled
  change is never extensive}.  Conversely, if subsystem energies are
entangled so the \mbox{total} energy distribution is finite and
uniform~\cite{macro-note}, \mbox{total} average energy causes maximal
distinct change, with each subsystem's average energy causing its share
\textit{regardless} of its Hamiltonian's locality: all change permitted
by~a subsystem's energy can be attributed to the subsystem
(\cf\cite{building-spacetime,er=epr,jacobson-qgr}).  This allows a
maximally distinct classical mechanics to be modeled as an
\textit{unentangled} product of maximally distinct subsystems,
\textit{if} we count the distinct change \textit{separately} in each
(\eg, see Appendix~\ref{appendix.field-theory}).

The near-maximal distinctness of random Hamiltonian evolution
\textit{without locality} suggests a conjecture on why a generic cosmic
evolution might be essentially {maximally} distinct: perhaps the
dynamics initially had no locality.  It seems plausible there was a
{\em Planck era} when dynamics was divorced from present day locality:
when all of the wavelengths observable in our expanding universe~today
extrapolate back to a sub-Planck-length
scale~\cite{early-universe,kempf0}.  Thus maximal distinctness may be a
realistic model for the evolution of large isolated systems today
\textit{if we take all entanglement into account}
(\cf\cite{zurek,bohm,lessons}).


\section{A classical quantum-field}
\label{appendix.field-theory}

In classical field theory, the principal ways to describe a flow field
are called {\em Lagrangian} and {\em Eulerian}~\cite{succi-book}.  In
the Lagrangian approach, we label individual parcels of the flow and
follow their motion---we follow the particles.  In the Eulerian
approach, we instead label fixed locations in space and watch what
flows through them.

A quantum lattice description of the computation of
Figure~\ref{fig.ssm} is purely Eulerian: a spatial array of unitary
logic gates~\cite{qc-gates} alternately swaps qubits between adjacent
lattice sites and updates the qubits that land at each site.  This does
not, however, realize the lattice dynamics as a special case of a
spatially-continuous quantum field in which a particle can start
anywhere and moves continuously.  We can achieve that by adopting a
Lagrangian description of the particle motion
(\cf\cite{succi-book,yepez-path,yepez-gauge,yepez-higgs}).


\begin{figure}[t]
  {\hfill\includegraphics[width=.45\columnwidth]{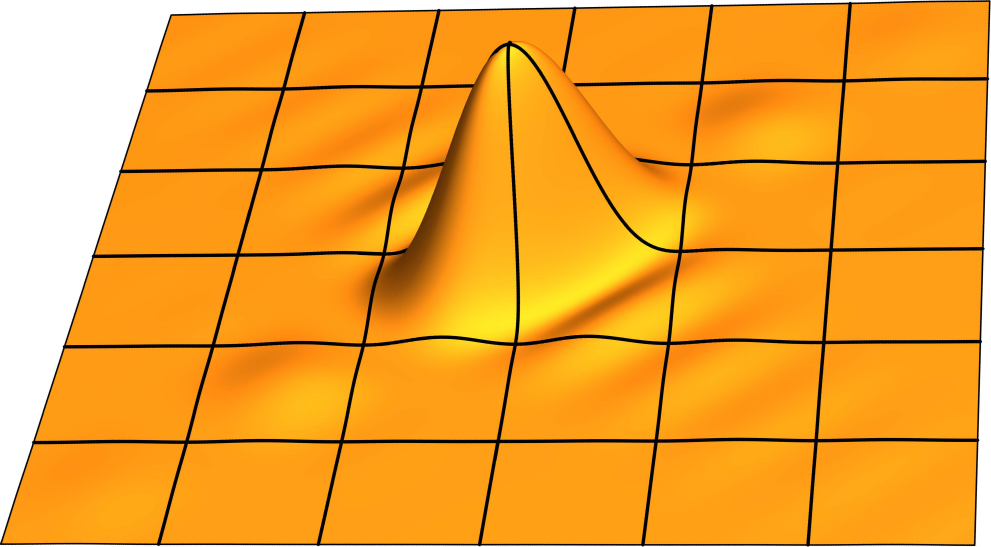}\hfill\hfill\includegraphics[width=.45\columnwidth]{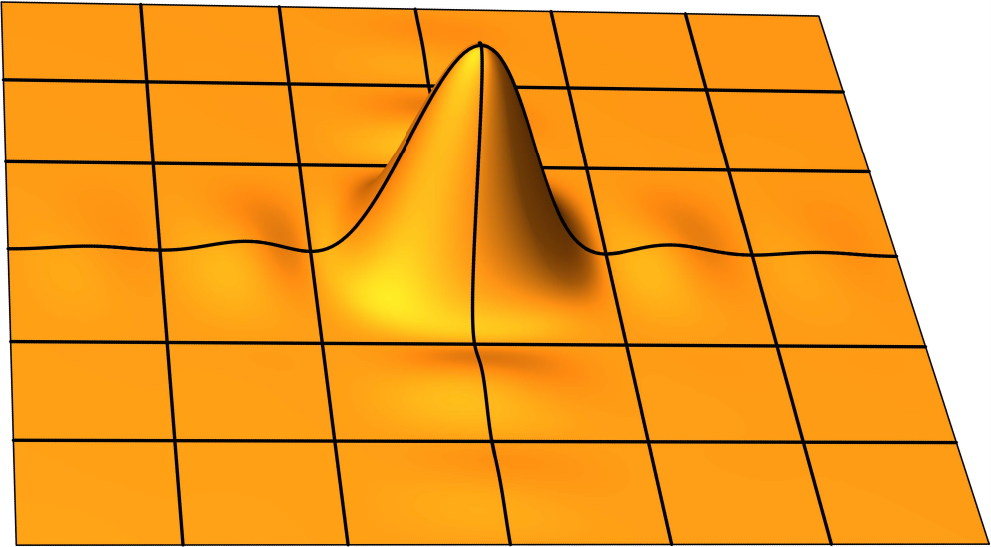}\hfill}
  \caption{ {\it A single-particle state $\av{\psi}^2\!$ centered at a
      grid point}, (a)$\ns$ moving diagonally, (b)$\ns$ moving along
    grid.  $\ns\psi$ is a product of two maximally distinct wavepackets
    centering~it in orthogonal directions. $\psi = 0$ at all but one
    grid point.  A particle centered {\em anywhere} is a superposition
    of states centered at grid points.}  \label{fig.single-particle}
  \end{figure}

Let $\ket{x,y,d}$ be the state of a {\em single particle} centered at
continuous coordinates $(x,y)$ moving in direction-$d$, where $d$ is
one of the directions of motion of the lattice gas model.  As a
superposition of position eigenstates, $\ket{x,y,d}$ is a product of
two maximally distinct $\sinc_{b,\infty}$ wavepackets of
Appendix~\ref{appendix.interpol} (\cf\cite{wannier}): one centers it in
direction-$d$ with spacing $\lambda_{d}$ between distinct positions and
all momenta in direction-$d$, the other centers it perpendicular to
that, spaced the distance between distinct parallel paths and with no
net momentum (Figure~\ref{fig.single-particle}). A
\textit{\mbox{movement} step} shifts $\ket{x,y,d}$ in direction-$d$ at
speed $v_{d}$.  This shift models relativistic frame motion of a
particle with average momentum $p= h\ps/\ps 2\lambda_{d}$ and energy
$c^2 p\ps/\ps v_{d}\,$.

Using integer grid coordinates, the full wavefunction at integer times
is a \textit{symmetrized sum of products} of single-particle states
centered at integer positions: relabelings of which identical-particle
is which all represent the same classical lattice configuration.  These
integer-time states form a {\em basis} for states at \textit{all
  times}, since any shift of a $\sinc_{b,\infty}$ wavepacket is a
superposition of wave\-packets at integer coordinates.  This defines
the {\em occupation number basis} of a bosonic field
(\cf\cite{marg-mech,emulation,thooft-determ}).

In the full dynamics, the Lagrangian movement step alternates with an
Eulerian {\em interaction step}, in which the particles centered at
each pair of integer coordinates are transformed using creation and
annihilation operators to implement the lattice-gas collision rule
separately at each integer position~\cite{collision-note}.  This rule
conserves the average momentum and energy defined by the movement
steps.

This unitary dynamics is similar to an effective field theory defined
on a lattice corresponding to a maximum energy. Here, though, it is the
finite {\em average} momentum and energy of particles that implies a
finite spatial and temporal distinctness.  When started from an
occupation number state with 0 or 1 particles for each \textit{position
  plus direction}, the field is isomorphic to a local qubit dynamics at
integer times.  We could \textit{enforce} this occupation constraint by
using an \textit{antisymmetrized} (fermionic) basis, making this model
a special case of a fermionic system. Then exact mappings between
fermionic and qubit basis states would depend on which sign we assign
to each anti\-symmetrized state, but for our ideal integer-time
classical simulations this sign is meaningless
(\cf\cite{abrams-lloyd,bravyi,qca-review}).

\vspace*{.1em}

\section{Effectively-discrete integration}
\label{appendix.analysis}

Maximally distinct evolution has finite energy bandwidth, allowing
discrete analysis to replace continuous
\cite{kempf1,kempf2,kempf3,kempf4,tsang,marg-mech,emulation}.
Integration becomes particularly simple.

\newcommand{\tk}{{\tau_{\sss\kappa}}}

\paragraph{\bf Integrals equal sums.}
Let $s(t)$ be a function with \textit{period} $T$ and a finite
frequency range, of \textit{bandwidth} $\W$.  Time $\tau=T/N$ between
distinct values is given by equality in \eqn{wt} and, as long as the
middle frequency $\nu_*=b/\tau$ of the finite range is not too far from
zero ($\av{\nu_*}\le\W/2$),
\begin{equation}\label{eq.int=sum}
\int_{\sss T} dt \, s(t)=
\raisebox{0pt}[0pt][8pt]{$\displaystyle\sum\nolimits$}_{\sss \ns T}
\tau \,s(n \tau)\;.
\end{equation}
That is, the integral and sum over one period are equal.  The identity
\eqn{int=sum} is just the time integral of \eqn{sampling}, since
$\int_{\sss -T/2}^{\sss T/2} dt\,\sinc_\bn (t/\tau-n)=\tau$ as long as
$\nu_*$ also gives exact periodicity.  This follows from \eqn{sincn},
integrating $t$ before summing $m$, and $bN\ns+\ps m$ an integer equal
to 0 for some~$m$.

We can compute the integral using more samples, but not less. In fact,
\eqn{int=sum} holds if we replace $\tau$ with $\tk = \tau/\kappa$ for
integer $\kappa\ge1$. This is because the sum must be the same if we
shift the origin of $t$, since that is true for the integral.  For
example, for $\tau\to\tau_{\sss 2}$, the sum turns into two equal
$\tau_{\sss 1}$-separated sums, each half-weighted.  In the limit
$T\to\infty$, $\kappa$ can be any real number $\ge1$ (see
\cite{krishnan}).

\paragraph{\bf Integrals of quantum averages equal sums.}
An even stronger result holds for quantum averages since both
$\sinc_\bn$ and $\sinc^*_\bn$ appear in the inner product, eliminating
any dependence on the center of a frequency range.  Given a
time-independent linear operator $\op{L}$ and a state $\ket{\psi}$ with
energy bandwidth $h\ps\W$ and period $T$ up to an overall phase (\ie,
$\av{\braket{\psi(t+T)}{\psi(t)}}=1$),
\begin{equation}\label{eq.inta=sum}
\int_{\sss T} \ns\ns dt\,\bra{\psi(t)}\op{L}\ket{\psi(t)}\ps=
\raisebox{0pt}[0pt][8pt]{$\displaystyle\sum\nolimits$}_{\sss T} \ns\tau\,\bra{\psi(n
  \tau)}\op{L}\ket{\psi(n\tau)}\;,
\end{equation}
where distinct separation $\tau=T/N$ is again given by~\eqn{wt}.  Since
\raisebox{0pt}[8pt][4pt]{$\int_{\sss -T/2}^{\sss T/2} \ns dt\ps
  \sinc_\bn^*(t/\tau
  -n^\prime)\,\sinc_\bn(t/\tau-n)=\tau\,\delta_{n^\prime n}$,}
\eqn{inta=sum} is just time integration with \eqn{sampling-time}
defining $\ket{\psi(t)}$, and again the same result holds for
$\tau\to\tk\,$.

More generally, \eqn{int=sum} extends to any periodic function with
finite conjugate bandwidth, \eqn{inta=sum} to any bandlimited
single-parameter unitary evolution.  They thus apply to isolated
classical systems modeled as maximally distinct for their energy and
momentum (see Appendix~\ref{appendix.large}).


\begin{thebibliography}{99}
\bibitem{noon-note} Let initial state
  $\ket{\psi}=(\ket{n,0}+\ket{0,n})/\sqrt{2}$, an equal superposition
  of $n$ photons in arm$_1$ of the interferometer with 0 in arm$_2$,
  and vice versa.  The free Hamiltonian for each arm$_i$ is
  $\op{H}_i=(\op{n}_i+\frac{1}{2})\,h\nu$, and if there is a time delay
  $\delta t$ in arm$_2$ relative to arm$_1$, the generator of distinct
  evolution is $\,\frac{\op{H}_1-\op{H_2}}{2}\,\delta t\,$ since the
  rest of the dynamics contributes just an overall phase: $H_1 t+ H_2
  (t-\delta t) = $ $\frac{\op{H}_1-\op{H_2}}{2}\ps\delta t +
  \frac{\op{H}_1+\op{H_2}}{2}\ps(2t-\delta t)$~\cite{yurke}.  Since
  $\ket{\psi}$ is an equal super\-position of eigenvalues $\pm n
  h\nu\ns\ns/2$ of $\frac{\op{H}_1-\op{H_2}}{2}$ it achieves \eqn{wt}:
  distinct states are $T/2\ps=\ps1/2\ps n\nu\ps=\ps
  h/2\ps(E-E_0)\raisebox{0pt}[8pt][0pt]{}$ apart.
\bibitem{mass-note} This lattice dynamics is consistent with an energy
  and momentum conserving classical mechanics, regardless of what speed
  $v\le c$ we attribute to the fastest particle.  Letting $v=c$ treats
  it as having no internal dynamics.
\bibitem{jacobson-private} Private communication from Ted Jacobson,
  suggesting a connection with the vacuum energy problem in GR: to
  gravitate, energy has to be able to cause distinct change.
\bibitem{bw1-note} This is a tight bound. Consider a portion with $N$
  distinct states with unit separations in time, except for one
  separation of length 2.  This has the same constraints \eqn{orthogt}
  as $N+1$ distinct states with unit separations.  Thus, from \eqn{wt},
  a periodic completion with period $T=N+1$ has $\W\ge N/T=N/(N+1)$,
  and the average separation within the portion is $\tau=N/(N-1)$,
  hence $\W\ps\tau\ge N^{\,2}/(N^{\,2}-1)>1$.
\bibitem{mand-note} From Lyapunov's inequality \cite{casella}, higher
  moments of a probability distribution bound lower ones, so second
  deviation from the mean $\W_2\le \W_{\sss \M}$ for $\M>2$. We get
  equality with two non-zero and equal probabilities; this state
  achieves the $\M=2$ bound $f_2\le \tau\,\W_2 \le \tau\,\W_{\sss \M}$.
\bibitem{sum-dist} In energy basis, total Hamiltonian is diagonal and
  each energy eigenvalue is the sum of $n$ corresponding \mbox{random}
  values of the $n$ local-Hamiltonian matrices. For large $n$ this
  random walk has a truncated normal PDF \cite{code}.
\bibitem{macro-note} We can always construct an evenly weighted range
  for~a dense energy spectrum by weighting each energy in the range by
  the interval to the next higher energy \cite[\S2.3]{max-speed}.
\end{thebibliography}
\end{document}